\documentclass[superscriptaddress,aps,amsmath,amssymb,showpacs,showkeys]{revtex4-2}
\usepackage[dvips]{graphicx}
\usepackage{times}
\usepackage{braket}
\usepackage{xcolor}
\usepackage{orcidlink}
\usepackage{hyperref}
\usepackage{booktabs} 
\usepackage{subfigure}
\usepackage{siunitx}
\hypersetup{
	colorlinks=true,
	urlcolor=magenta,
	linkcolor=red,
	citecolor=blue
}

\begin{document}
\title{Ageing and Quenching: Influence of Galaxy Environment and Nuclear Activity in Transition Stage}
	
	\author{Pius Privatus\orcidlink{0000-0002-6981-717X}}
	\email[Email: ]{privatuspius08@gmail.com}
	\affiliation{Department of Physics, Dibrugarh University, Dibrugarh 786004, Assam, India}
	\affiliation{Department of Natural Sciences, Mbeya University of Science and Technology, Iyunga 53119, Mbeya, Tanzania}
	
	\author{Umananda Dev Goswami\footnote{Corresponding author}\orcidlink{0000-0003-0012-7549}}
	\email[Email: ]{umananda@dibru.ac.in}
	\affiliation{Department of Physics, Dibrugarh University, Dibrugarh 786004, Assam, India}

	%\date{}
	\begin{abstract}
This study aims to investigate whether the environment and the nuclear activity 
of a particular galaxy influence the ageing and quenching at the transition 
stage of the galaxy evolution using the volume-limited sample constructed from 
the twelve release of the Sloan Digital Sky Survey. To this end, the galaxies 
were classified into isolated and non-isolated environments and then each 
subsample was further classified according to their nuclear activity using 
the WHAN diagnostic diagram, and ageing diagram to obtain ageing and quenching 
galaxies. The ageing and quenching galaxies at the transition stage were 
selected for the rest of the analysis. Using the star formation rate and the 
$u-r$ colour-stellar mass diagrams, the study revealed a significant change of 
$0.03$ dex in slope and $0.30$ dex in intercept for ageing galaxies and an 
insignificant change of $0.02$ dex in slope and $0.12$ dex in intercept of the
star formation main sequence between isolated and non-isolated quenching 
galaxies. Further, a more significant change in the number of ageing galaxies 
above, within and below the main sequence and the green valley was observed. 
On the other hand, an insignificant change in the number of quenching galaxies 
above, within and below the main sequence and the green valley was observed. 
The study concludes that ageing depends on the environment and the 
dependence is influenced by the nuclear activity of a particular galaxy while 
quenching does not depend on the environment and this independence is not 
influenced by the nuclear activity.
	\end{abstract}
	
	%\pacs{}
	
\maketitle    

\section{Introduction}
The processes of galaxy evolution and the end of star formation within the 
galaxy remain in hot debate to date, with unsolved issues including 
the role of the environment and the nature of the feedback mechanism. This is 
due to the 
complex nature of the physical processes underlying this field, despite the 
fact that the relation between the star formation history (SFH) of galaxies 
and their physical properties such as stellar mass, environment, morphology 
and chemical composition has been extensively studied during the last few 
decades \cite{kauffmann2004environmental,tremonti2004origin,peng2010mass}. 
The distribution of most of the galaxy's physical properties including  
star formation rate (SFR), stellar mass (M$\star$) and colour have been 
proven to be bimodal, e.g. in Refs.~\cite{peng2010mass,brinchmann2004physical} 
the galaxies are classified into star forming with abundant gas reservoirs and 
passive that correspond to gas-poor
\cite{gonccalves2012quenching,moustakas2013primus}.  Due to such 
bimodal distributions of properties, the galaxies are divided with respect to 
the main sequence (MS, having a tight correlation between SFR and M$\star$) 
into within (normal star forming), above (starburst) and below (passive) the
MS, where the starburst are galaxies that undergo excess star 
formation activity and the passive have little star formation activity 
\cite{elbaz2007reversal,speagle2014highly,leslie2015quenching,
daddi2007multiwavelength,yuan2010role,rich2011galaxy,
schawinski2007observational, whitaker2012star,shimizu2015decreased}. The 
star forming was observed to be in the blue cloud (BC), while the retired 
galaxies were in red sequence (RS). The galaxies in intermediate were defined 
to be in the green valley (GV) in colour against M$\star$ diagrams 
\cite{faber2007galaxy,hickox2014black,schawinski2014green}.
	
In recent years different internal and external (environmental) processes have 
been proposed as the main reason for the transition from the BC to the RS 
mostly characterised by the decrease 
in SFR referred to as quenching and ageing. Quenching has been 
defined as a mechanism able to terminate the star formation process in a 
galaxy by an agent (e.g.~negative feedback from Active Galactic Nuclei (AGN) 
or supernovae winds) which causes either the removal or heating of the cold 
gases necessary for star formation, while the term ageing denotes the 
continuous evolution of a galaxy through a secular process driven by the 
life cycle of its stars \cite{corcho2023ageinga,corcho2023ageingb}. For the 
internal processes the galaxy may 
quench due to internally triggered quenching mechanisms such as negative 
feedback from AGN, supernovae winds or the 
stabilisation of the gas against fragmentation 
\cite{fitts2017fire,gensior2020heart}. On the other hand, external processes 
triggered quenching may be 
due to environmentally driven processes such as ram pressure stripping able 
to remove the gas reservoir, strangulation or starvation (leading to the 
suppression of gas infall or galaxy interactions) 
\cite{brown2017cold,cortese2021dawes}. These processes are observed 
to slow down the rate of star formation but also enhance the star formation 
in a short time scale \cite{thorp2022almaquest}. Quenching has been categorised into fast and slow quenching 
depending on the duration of time spent to cross an arbitrary boundary in 
the M$\star$ versus specific star formation rate 
(SSFR = SFR/M$\star$) plane and colour-magnitude diagram 
\cite{akins2022quenching,tacchella2022fast,suess2022studying}. The fast 
quenching involves the termination of SFR in a short time scale that is less 
than 1 Gyr and slow quenching involves the termination of SFR with a time 
scale comparable to the age of the universe. In this regard ageing 
is different but difficult to distinguish from slow quenching, they 
significantly differ in the fact that quenching involve agents and active 
suppression of star formation while ageing does not, more important to keep in 
mind is that both processes results in the transition of galaxies from BC to 
RS \cite{corcho2023ageinga,corcho2023ageingb}.
	
Ref.~\cite{belfiore2018sdss} studied the radial profiles 
of emission lines from galaxies to quantify 
the $H\alpha$ emission line strength in a galaxy's spectrum relative to its 
continuum emission ($H\alpha$ equivalent width, denoted as EW($H\alpha$)) and 
SSFR derived from spatially resolved Mapping Nearby Galaxies at Apache 
Point Observatory (MaNGA) survey as detailed in 
Ref.\ \cite{abdurro2022seventeenth}, 
to gain insight into the physical mechanisms that suppress the star formation, 
and observed that the responsible quenching mechanism appears to affect the 
entire galaxy. Ref.\ \cite{erfanianfar2016non} obtained that morphology and 
environment have a combined role in slowing down the star formation activities 
in galaxies. Furthermore, they observed that a long-timescale environmental 
effect appears at low redshift. It needs to be mentioned that our 
current study specifically focuses on the effects of environment and nuclear 
activity on the galaxies' star formation activities, keeping the morphology 
effect as a future prospect of a detailed study.
  Ref.\ \cite{lang2014bulge} suggested that the 
decrease of SFR is mainly due to internal process and is linked with 
bulge growth. However, the existence of the relation between the morphology and 
density may lead to a change in the relation of SFR, stellar mass and 
the environment where a particular galaxy resides. The study by 
Ref.\ \cite{bluck2020galactic}, presenting the analysis of star formation 
and quenching in the SDSS MaNGA survey as detailed in 
Ref.\ \cite{abdurro2022seventeenth}, utilising over 5 million spaxels from 
$\sim 3500$ local galaxies observed that the sudden decrease of SFR affect 
the whole galaxy but star formation occurs in a small localised scale within 
the galaxy. On the other hand Ref.\ \cite{bluck2020galactic} observed that 
quenching is global while star formation is governed by local processes within 
each pixel. All these studies aim to discern the mechanism or combination of 
mechanisms that lead to the quenching of star formation processes in galaxies, 
thereby influencing their evolution. 

In a simplified representation, as already stated the processes 
guiding the galaxy transformations can be broadly categorised as internal and 
external mechanisms. Processes like negative feedback from AGN, supernovae 
winds act within the galaxy and are classified under internal mechanisms, while 
processes like ram-pressure stripping and galaxy mergers originate from the
external mechanisms. The study by Ref.\ \cite{speagle2014highly}, pointed out 
that correlation between SFR and M$\star$ decreases with redshift 
($\propto (1 + z)^\gamma$, where $\gamma$ ranges from $1.9$ to $3.7$) rather 
than being driven solely by stochastic events like major mergers or 
starbursts \cite{leslie2020vla,thorne2021deep,leja2022new}. 
On the other hand, the study by Ref.\ \cite{croton2006many} using the 
semi-analytic model of galaxy formation, observed that the AGN feedback is 
the primary mechanism affecting the quenching. Although this was observed to 
affect the massive galaxies (with M$\star \geq10^{11}$ M$\odot$), hence there 
is a lack of observed feedback effects for the majority of the star forming 
galaxies \cite{rosario2012mean}. However, Ref.\ \cite{oemler2017star} using a 
basic model for disk evolution, demonstrated that the evolution of galaxies 
away from the main sequence can be attributed to the depletion of gas due to 
star 
formation after a cut-off of gas inflow. This model was based on the observed 
dependence of star formation on gas content in local galaxies and assuming 
simple histories of cold gas inflow. Again Ref.\ \cite{oemler2017star} using 
the MS as the tracer of the factors responsible for quenching 
further obtained that galaxies classified as MS, quiescent (with reduced star formation), or 
passive (with no ongoing star formation) exhibit varying fractions on mass and environment. 
The MS fractions 
decrease with increasing mass and density, while the quiescent and passive 
fractions rise. Due to this uncertainty, the ongoing debate revolves around 
the extent to which each of these scenarios influences the shape of the 
relationship.

By examining the physical characteristics of MaNGA as detailed in 
Ref.\ \cite{abdurro2022seventeenth} and the Sydney-Australian Astronomical 
Observatory (AAO) Multi-object Integral field spectrograph (SAMI) Galaxy Survey as detailed in  
Refs.\ \cite{croom2012sydney,bryant2015sami}, Ref.~\cite{corcho2023ageingb} 
found that the ageing population is made up of a heterogeneous mixture of 
galaxies, primarily late-type systems (e.g.~spiral and irregular galaxies), since 
they start with high SFR, as they becomes older their SFR decreases results to the 
transition from star-forming (high SFR) to quiescent states (reduced SFR) with 
a range of physical features. Authors pointed out that the retired 
(galaxy systems that have ceased star formation) were formerly 
ageing or quenched. These galaxies are found across different masses 
and environments ranging from low-mass (M$\star<10^{11}$ M$\odot$), 
low-density ($\Sigma_{5}<1$ Mpc$^{-2}$) regions to high-mass 
(M$\star\geq10^{11}$ M$\odot$), high-density ($\Sigma_{5}\geq 1$ Mpc$^{-2}$) 
ones, where $\Sigma_{5}$ is the surface density of galaxies to the 5th nearest 
neighbour given by $\Sigma_5 = {5}/{\pi d_5^2}$. Here $d_5$ is the projected 
distance to the 5th nearest neighbouring galaxy, measured in Mpc.
In their analysis they shows that 
distinguishing between retired and recently quenched galaxies are 
important to constrain the mechanisms driving galaxy evolution. 
They further left two questions viz., whether galaxies evolve due to 
their initial conditions at formation such as chemical composition, the angular 
momentum, mass of the galaxy and the initial density of environment (nature)? 
Or the external quenching mechanism such as the environment, galaxy 
interactions and AGN feedback (nurture)? Distinguishing between these factors 
is very important in understanding galaxy transformation.

As the follow-up of  the work by Ref.\ \cite{corcho2023ageingb}, this  study 
aims to investigate whether the environment influences ageing and quenching, 
and how their relationships with the environment are affected by nuclear 
activity. This study uses the friends-of-friend method from 
Ref.~\cite{tempel2017merging}, to assign ageing and quenching galaxies in a 
transition stage into systems of isolated (with no neighbour) and non-isolated 
(with at least one neighbour) environment and then to compare their equations 
of the main sequence, the colour obtained from $u$- and $r$-band 
luminosities ($u-r$ colour) and stellar mass diagrams.
	
Our paper is organized as follows. In the next section, we explain the source 
of data and the method of getting samples. In Section \ref{secIII} we explain 
the methodology used in this study. The Section \ref{secIV} is dedicated to 
presenting the results. In Section \ref{secV} the results are discussed. 
Section \ref{secVI} presents the summary and conclusion. Cosmological 
constants used in this work are adopted from 
Ref.\ \cite{collaborartion2016planck}, wherein the dark energy density 
parameter $\Omega_{\Lambda}=0.692$, Hubble constant 
$H_0=67.8$ km s$^{-1}$ Mpc$^{-1}$ and the matter density parameter 
$\Omega_{m}=0.308$ are recorded.	
\section{Data} \label{secII}
\subsection{The SDSS main sample}\label{sdss}
In this research work the catalogue data extracted from the flux-limited 
sample of twelve releases of Sloan Digital Sky Survey (SDSS DR12) as detailed 
in Ref.\ \cite{eisenstein2011sdss,alam2015eleventh} are used. The main galaxy 
sample was selected from the main contiguous area of the survey based on the 
methods outlined in Ref.\ \cite{tempel2017merging}. Galaxy data were 
downloaded from the SDSS Catalogue Archive Server (CAS). The objects with the 
spectroscopic class GALAXY or QSO as detailed in 
Ref.\ \cite{alam2015eleventh}, were selected as suggested by the SDSS team. We 
then filtered out galaxies with the galactic-extinction-correction based on 
Ref.\ \cite{schlegel1998maps} where Petrosian r-band magnitude fainter than 
$17.77$ are rejected keeping in mind that the SDSS is incomplete at fainter 
magnitudes \cite{strauss2002spectroscopic}. After correcting the redshift for 
the motion with respect to the cosmic microwave background (CMB), using the 
simplified formula $z_{\text{CMB}} = z_{\text{obs}} - v_p/c$, where $v_p$ is 
a velocity of motion along the line of sight relative to the CMB, the upper 
distance limit at $z = 0.2$ was set. The final data set contains $584449$ 
galaxies.
\subsection{The volume-limited samples}\label{vl}
As already stated the SDSS main data are flux-limited, one of the 
disadvantages of using the flux-limited sample is that only the 
luminous objects have a chance to be observed at large distances. Due to this
main reason, the volume-limited samples are desired. Hence we constructed a 
volume-limited sample for uniformity. Due to the peculiar velocities of 
galaxies in groups, the measured redshift (recession velocity) does not give 
an accurate distance to a galaxy located in a group or cluster, the apparent 
magnitude was transformed into absolute magnitude using the relation:
\begin{equation}
M_{r}= m_r-25-5\log_{10} (d_L)-K,
\label{eqR} 
\end{equation}
where $d_L$ is the luminosity distance, $M_{r}$ and $m_{r}$ are r-band 
absolute and apparent magnitudes respectively, and K is the k+e-correction. 
The k-corrections were calculated with the KCORRECT ($v4\_2$) algorithm 
\cite{blanton2007k}. The evolution corrections were estimated using the 
luminosity evolution model of $K_e = c*z$, where $c = -4.22, -2.04, -1.62, -1.61, -0.76$ for the 
u,g,r,i,z-filters, respectively \cite{blanton2003galaxy}. The magnitudes corresponding to the rest 
frame (at the redshift $z = 0$) and evolution correction were estimated 
similarly by assuming a distance-independent luminosity function 
\cite{tempel2012groups,tempel2014flux}. According to 
Ref.\ \cite{ball2006bivariatec}, the Schechter function's 
\cite{schechter1976analytic} typical magnitude $M_r^{*}$ is around $-20.5$ 
mag. The physical properties of galaxies have been shown to undergo an abrupt 
transition at the typical magnitude $M_r^{*}$. Clustering depends 
weakly on the environment for galaxies fainter than $M_r^{*}$ but is stronger 
for brighter galaxies \cite{deng2012some}.
We have constructed the volume-limited samples above the characteristic 
magnitude $M_r^{*}$ by calculating the effective maximum distance using the 
relation as given by 
\begin{equation}
d_{\text{max}} = 10^{\left(m_{r\text{lim}} - M_{r\text{min}} + 5\right)/5} \times 10^{-6} (\text{Mpc}),
\label{dmax}
\end{equation}
where $m_{r\text{lim}}=17.17$ mag, $M_{r\text{min}} =-20.5$ mag.
Using the $d_{\text{max}}$ values and the luminosity restrictions we 
constructed the volume-limited main galaxy sample containing $136274$ 
galaxies within $ -22.5\leq M_r \leq -20.5$ (mag). 
	
\section{Methodology}
\label{secIII}
\subsection{Isolated and non-isolated environments}\label{env}
Using the volume-limited sample defined in the previous section we assigned the 
galaxies into isolated and non-isolated sub-samples, compiled using the 
friend-of-friend (FoF) method with a variable linking length. The essence of 
this approach lies in the division of the sample into distinct systems 
through an objective and automated process. This involves creating spheres 
with a linking length ($R$) around each sample point (galaxies). 
The linking length was adjusted using the arctangent law as given by 
\begin{equation}
R_{LL}(z) = R_{LL,0} \left[1 + a \arctan\left(\frac{z}{z_\star}\right)\right],
\label{eq1} 
\end{equation}
where $R_{LL}(z)$ is the linking length used to create a sphere at a specific 
redshift, $R_{LL,0}$ is the linking length at $z = 0$, $a$ and ${z_\star}$ are 
free parameters. We obtained the linking length (which defines a sphere used 
to group galaxies) by fitting the arctangent law defined by equation 
\eqref{eq1} to the observational data (nearest neighbour distance with 
redshift). The best-fit values for the parameters were: $R_{LL,0}=0.34$ Mpc, 
$a=1.4$ and 
${z_\star}=0.09$ \cite{tempel2017merging}. If there are other galaxies within the sphere, 
they are considered as the parts of the same system and referred to as ``friends''.  
Subsequently, additional spheres are drawn around these newly identified 
neighbours, and the process continues with the principle that ``any friend of
my friend is my friend''. This iterative procedure persists until no new 
neighbours or ``friends'' can be added. At that point, the process concludes, 
and a system is defined. The galaxies in the system with no neighbour 
($N\!gal = 1 $ ) are isolated, while the galaxies with more than one 
neighbour ($N\!gal \geq 2$) are non-isolated. Consequently, each system 
comprises an isolated galaxy or a non-isolated galaxy that shares at least one 
neighbour within a distance not exceeding $R$. A total of $58032\,(42.59\%)$ 
isolated and $78233\,(57.41\%)$ non-isolated galaxies were obtained.

\subsection{Galaxy properties}\label{pro}
The stellar masses used in this study were obtained from the MPA-JHU team and 
calculated from the Bayesian approach as detailed in 
Ref.\ \cite{kauffmann2004environmental}. Stellar mass calculation within the 
SDSS spectroscopic fiber aperture relies on the fiber magnitudes, whereas the 
total stellar mass is determined using the model magnitudes. 
	
The MPA-JHU total SFR used in this study was derived from the MPA-JHU database 
and estimated using the methods of 
Refs.\ \cite{brinchmann2004physical,tremonti2004origin} with adjustments made 
for non-star forming galaxies. The MPA-JHU team uses the H$\alpha$ calibration 
outlined in Ref.\ \cite{kennicutt1998star} to determine the SFR for galaxies 
classed as star forming. In contrast to the approach taken by 
Ref.\ \cite{brinchmann2004physical}, the MPA-JHU team applied aperture 
corrections for SFR by fitting the photometric data from the outer regions of 
the galaxies. Specifically, for SFR computation, 
Ref.\ \cite{brinchmann2004physical} outlined the calculation within the galaxy 
fiber aperture. It is important to keep in mind that the region beyond 
the fiber SFR is estimated using the methods of  Ref.\ \cite{salim2015mass}.
Furthermore, for the case of AGN and weak emission line galaxies, SFR was 
determined using photometry. Keeping in mind that in the case of non-star 
forming galaxies, the ionisation originates from other sources, such as 
rejuvenation in the outer regions, post-AGB stars, or ionisation from AGN. As 
such, the SFR based on $H\alpha$ for non-star forming galaxies need to be 
regarded as a maximum value \cite{sanchez2018sdss}. 

\subsection{Galaxy classification} 
\begin{figure}[t!]
 \subfigure{
  \includegraphics[width=0.45\linewidth]{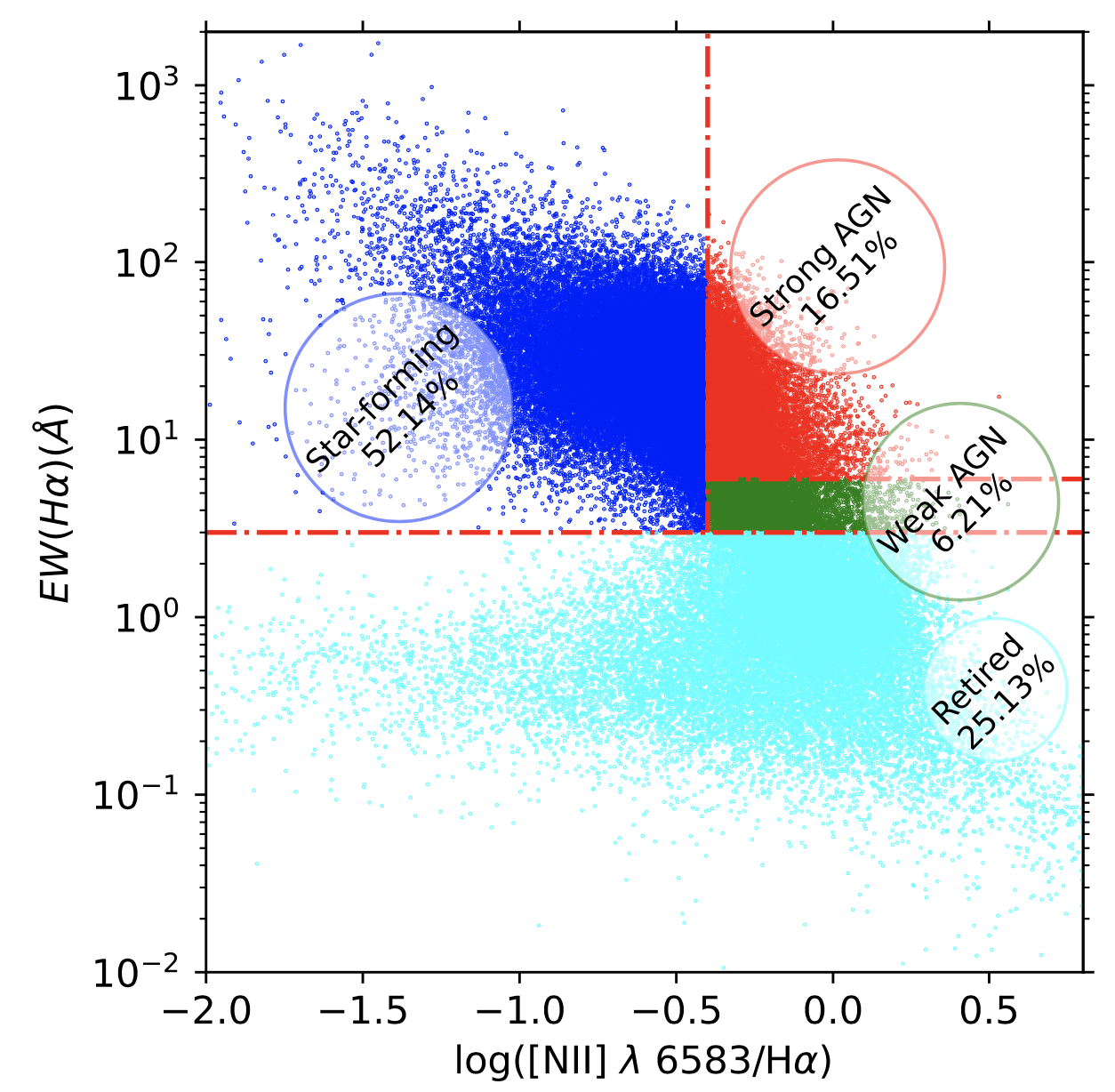}
    \label{Lisolated}
		}
		\hspace{0.3cm}
		\subfigure{
		 \includegraphics[width=0.45\linewidth]{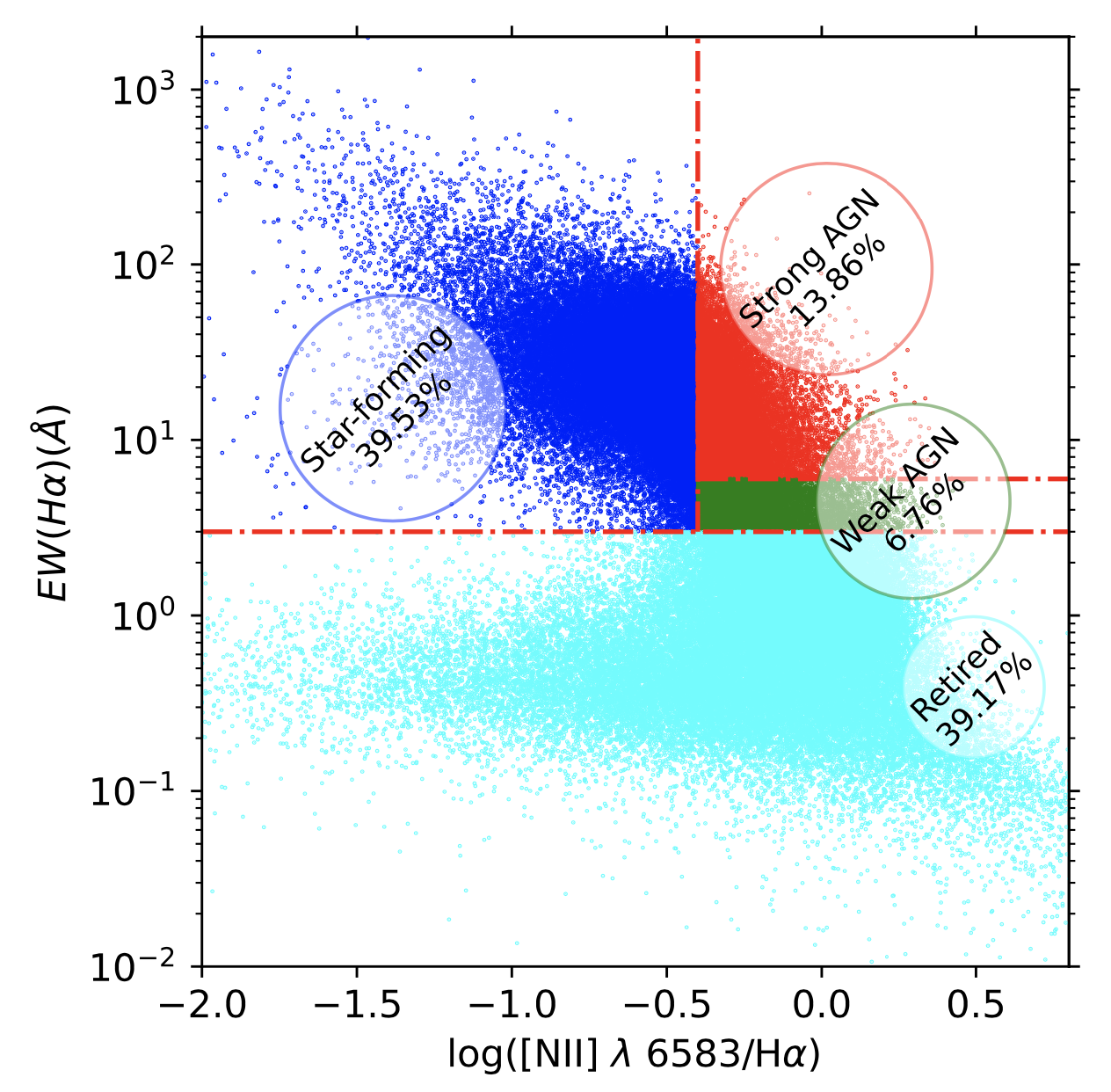}
			\label{Lnonisolated}
		}
		\vspace{-0.2cm}
\caption{The WHAN diagrams for the volume-limited main galaxy sample for 
isolated (left panel) and non-isolated (right panel) cases.}
	\label{WHAN}
\end{figure}

Intending to study where nuclear activity affects the dependence of ageing and 
quenching on environment, we classified the galaxies based on nuclear activity 
using the $W_{\text{H} \alpha}$ versus $[\text{NII}]/\text{H}\alpha$ (WHAN) 
diagram as detailed in Ref.\ \cite{cid2011comprehensive}, where 
$W_{\text{H}\alpha}$ is the $H\alpha$ equivalent width, and 
$[\text{NII}]/\text{H}\alpha$ is the ratio of the [NII] emission line to the 
$\text{H}\alpha$ line. From this diagram we can have following four 
inequalities \cite{cid2011comprehensive}:
	\begin{align}
		\log \left( \frac{[\text{N II}]}{\text{H}\alpha} \right) < -0.4 \quad \text{and} \quad W_{\text{H}\alpha} > 3 \, \text{\AA},
		\label{sf}\\[8pt]
		\log \left( \frac{[\text{N II}]}{\text{H}\alpha} \right) > -0.4 \quad \text{and} \quad W_{\text{H}\alpha} > 6 \, \text{\AA},
		\label{sAGN}\\[8pt]
		\log \left( \frac{[\text{N II}]}{\text{H}\alpha} \right) > -0.4 \quad \text{and} \quad 3 \, \text{\AA} < W_{\text{H}\alpha} < 6 \, \text{\AA},
		\label{wAGN}\\[8pt]
		W_{\text{H}\alpha} < 3 \, \text{\AA}.
		\label{RGs}
	\end{align}
These inequalities \eqref{sf}, \eqref{sAGN},
\eqref{wAGN}, and \eqref{RGs} represent the pure star-forming (SF)
galaxies, strong AGN, weak AGN and retired galaxies (RGs) respectively
\cite{cid2011comprehensive}. The use of the WHAN diagram is based on the
fact that the usual traditional diagnostic diagrams detailed in
Refs.\ \cite{kewley2001theoretical,kauffmann2003host,kewley2006host, 
schawinski2007observational}, may introduce bias since it is well-known that
shock ionisation and AGNs could cover almost any region between the right-low
end of the loci usually assigned to SF regions, up to the top-right end of
the diagram. On the other hand low-ionisation nuclear emission-line region 
(LINER) may contain possible multiple ionising sources 
\cite{cid2011comprehensive,singh2013nature,sanchez2020spatially,sanchez2021local}. 
The distributions of 
galaxies are shown in Fig.~\ref{WHAN} wherein the SF galaxies are shown by 
blue colour, strong AGN by red colour, weak AGN by green colour and retired 
galaxies by cyan colour. The following numbers was obtained for isolated 
galaxies:  $30261$ ($\sim 52\%$) SF galaxies; $9582$ ($\sim 17\%$) strong AGN; 
$3604$ ($\sim 6\%$) weak AGN and $14585$ ($\sim 25\%$) retired, while for 
the non-isolated sample: $31137$ ($\sim 40\%$) SF galaxies; $10921$ ($\sim 14\%$) 
strong AGN; $5320$ ($\sim 7\%$) weak AGN and $30855$ ($\sim 39\%$) retired. 
	
\begin{figure}[t!]
   \subfigure{
	\includegraphics[width=0.45\linewidth]{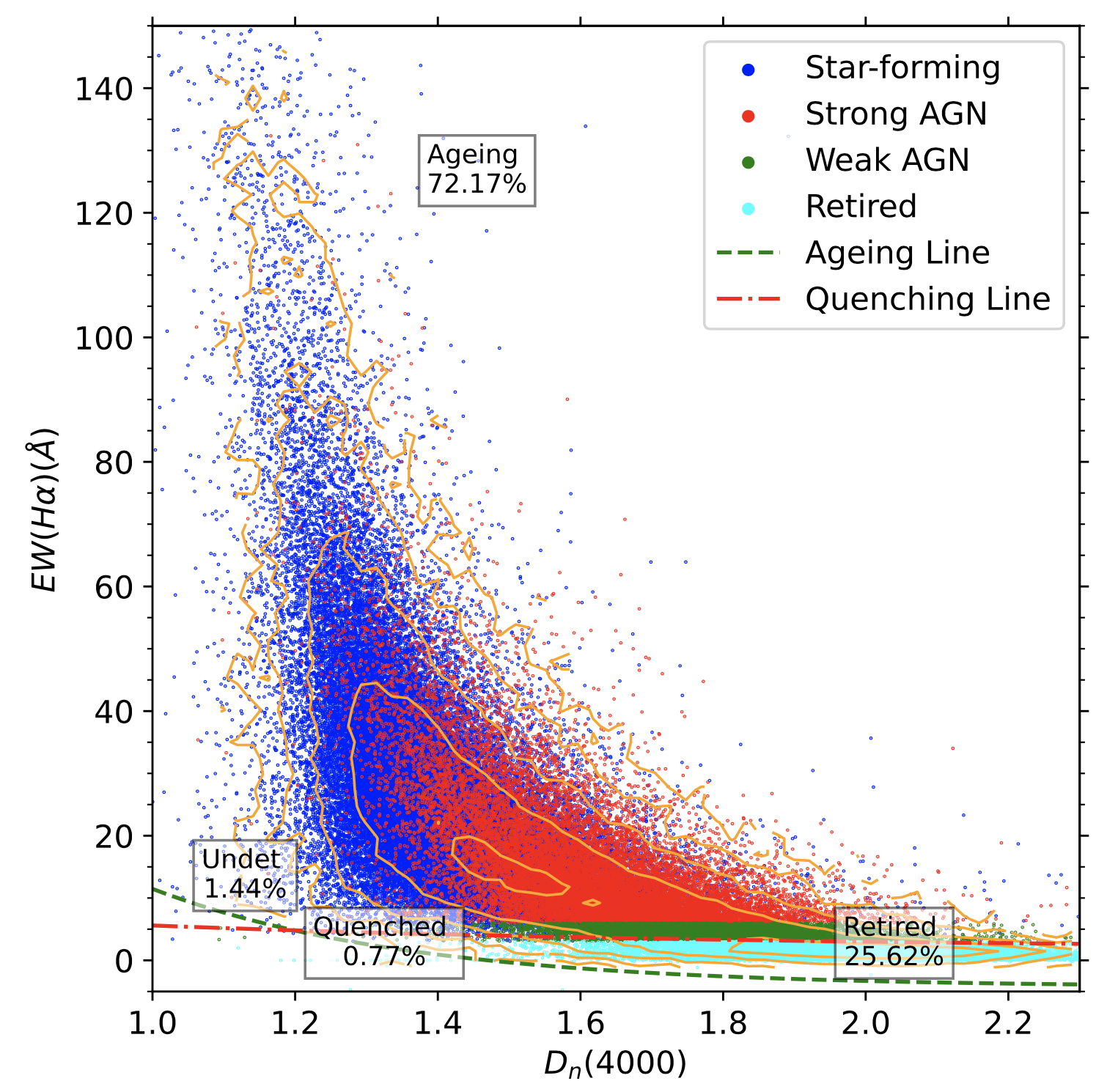}
        	\label{isolatedAD}
		}
		\hspace{0.3cm}
		\subfigure{
			\includegraphics[width=0.45\linewidth]{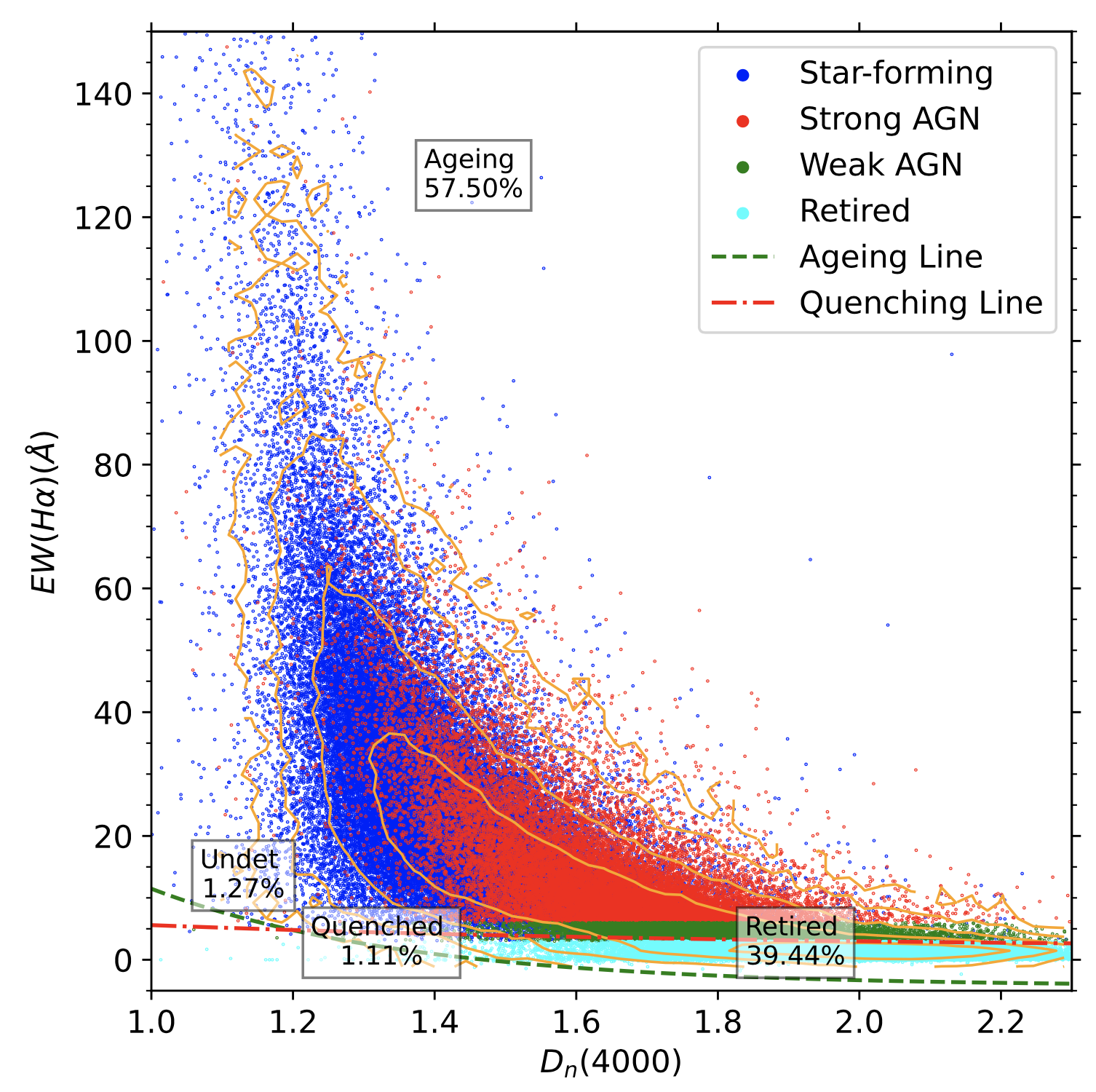}
			\label{nonisolatedAD}
		}
		\vspace{-0.2cm}
\caption{Ageing diagrams (ADs) for the volume-limited main galaxy sample for 
isolated (left panel) and non-isolated (right panel) cases.}
		\label{AD}
\end{figure}

We classified the galaxies into ageing and quenching systems using the ageing 
diagrams (ADs) as detailed in Ref.\ \cite{corcho2023ageinga,corcho2023ageingb}, 
given by Eqs.~\eqref{AG} (ageing line) and ~\eqref{QG} (quenching line )respectively. The galaxies located 
above ageing and quenching lines are classified as ageing galaxies (AGs) while the galaxies 
below  ageing and quenching line are quenching galaxies (QGs). The galaxies below the quenching line 
and above the ageing line are retired while the ones above the quenching line 
and below the ageing line are undetermined. The ADs are shown in Fig.~\ref{AD}, 
whereby for isolated a total number of 41886 ($\sim 72 \%$), 447 ($\sim 1\%$),  
14869 ($\sim 26  \%$), and 836 ($ \sim 1\%$) were obtained respectively for AGs, 
QGs, retired and undetermined galaxies, while for non-isolated a total number 
of 46862 ($\sim 58\%$), 876 ($\sim 1 \%$), 31066 ($\sim 40 \%$), and 1001 ($\sim 1 \%$) 
were obtained for AGs, QGs, retired and undetermined galaxies. It is very 
important to keep in mind that retired galaxies originate from both quenching 
and ageing processes, whereby the galaxies transit from blue cloud to red 
sequence. This transition is measured in terms of the colour index 
parameter $(g-r)$  which is the difference between $g$ (green)- and $r$ 
(red)-band luminosities. The lower values of the colour index ($g-r \leq 0.5$) 
indicate that the galaxy is in a blue cloud, while its higher values 
($g-r \geq 0.7$) indicate that the galaxy is in a red sequence. Since in this 
study our target is the galaxies in transition stage it is necessary to use 
the criteria between blue cloud and red sequence given by 
$0.5 < (g - r) < 0.7$, which implies that the galaxies are neither in blue 
cloud nor in red sequence. Keeping in mind that the associated uncertainties in 
measurements (e.g.~sky background and noise) of $g$ (green)- and $r$ 
(red)-band luminosities may affect the classification, we minimized the 
uncertainty in the $g$-band using the same method used for the  $r$-band 
luminosity through equation \eqref{eqR} as explained under Section \ref{vl}.
Moreover, we used the luminous galaxies to minimize the effect of noise.
A total of $16929$ and $16523$ isolated and 
non-isolated ageing galaxies in transition (AGT) was obtained while a total of 
$133$ and $286$ isolated and non-isolated quenched galaxies in the transition 
stage (QGT) were obtained. We will use the AGT and QGT samples in the next 
sections unless otherwise stated.
	
\begin{align}
\text{Ageing: EW(H}\alpha\text{)} / \text{\AA} &= 250.0 \cdot 10^{-1.2 \cdot D_n(4000) - 4.3}, \label{AG}\\[8pt]
\text{Quenched: EW(H}\alpha\text{)} / \text{\AA} &= -12.0 \cdot 10^{-0.5 \cdot D_n(4000) + 1.8}. \label{QG}
\end{align}
	
\section{Results}
\label{secIV}
\subsection{Ageing} \label{subsecIVa}
\begin{figure}[t!]
	\subfigure{
		\includegraphics[width=0.45\linewidth]{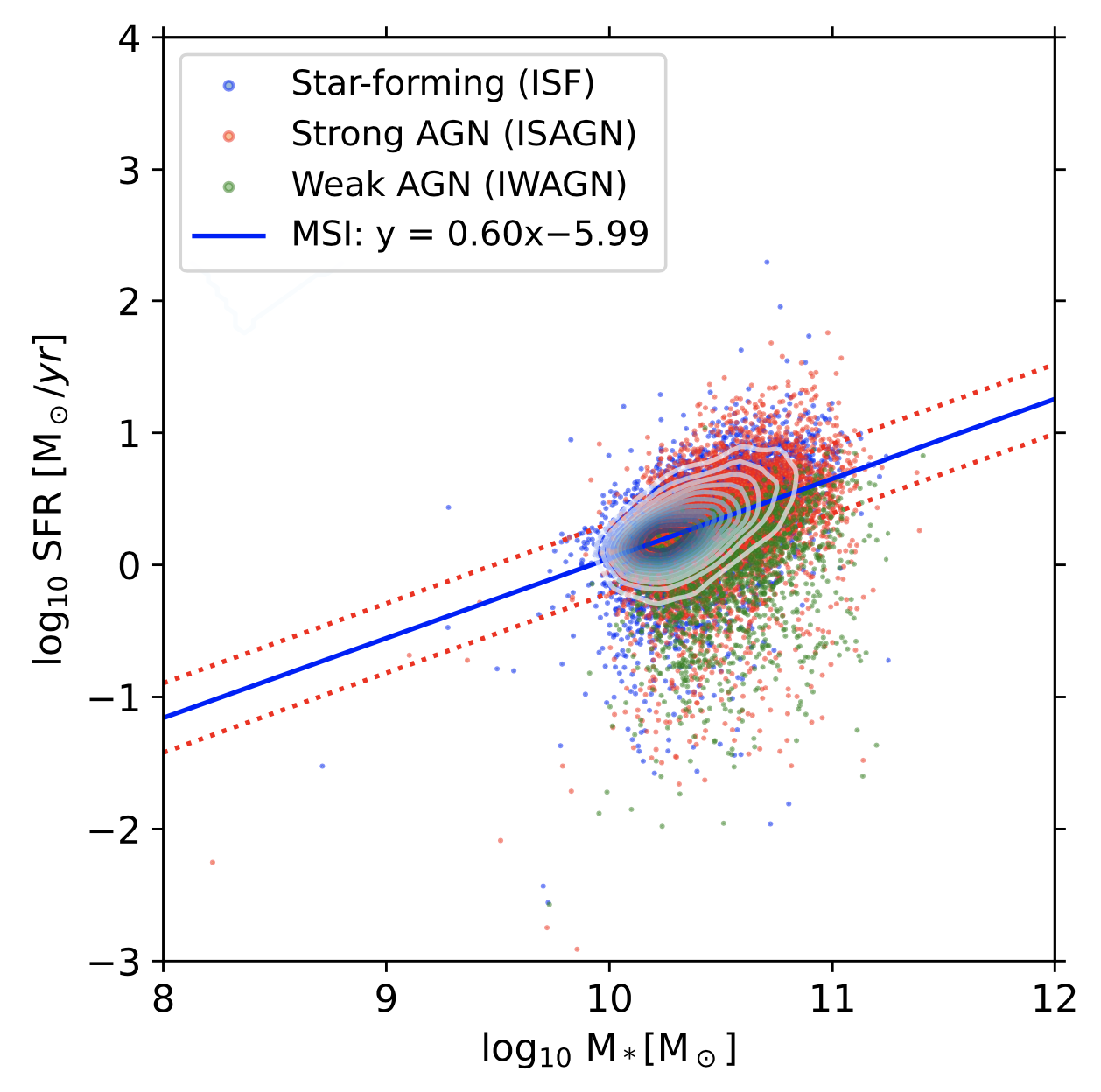}
			\label{MSAI}
		}
		\hspace{0.1cm}
		\subfigure{
			\includegraphics[width=0.45\linewidth]{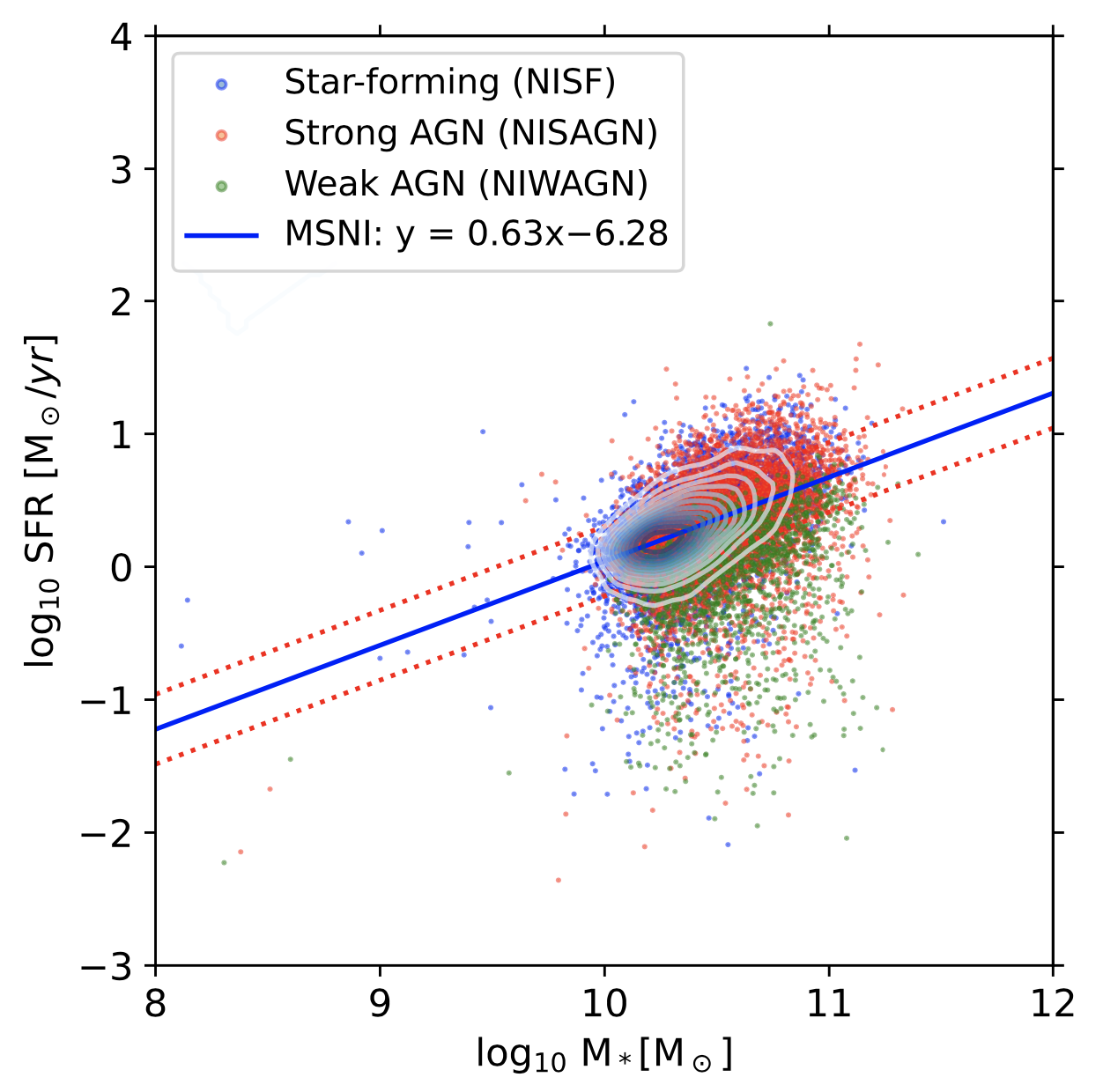}
			\label{MSANI}
		}
		\hspace{0.1cm}
		\subfigure{
			\includegraphics[width=0.45\linewidth]{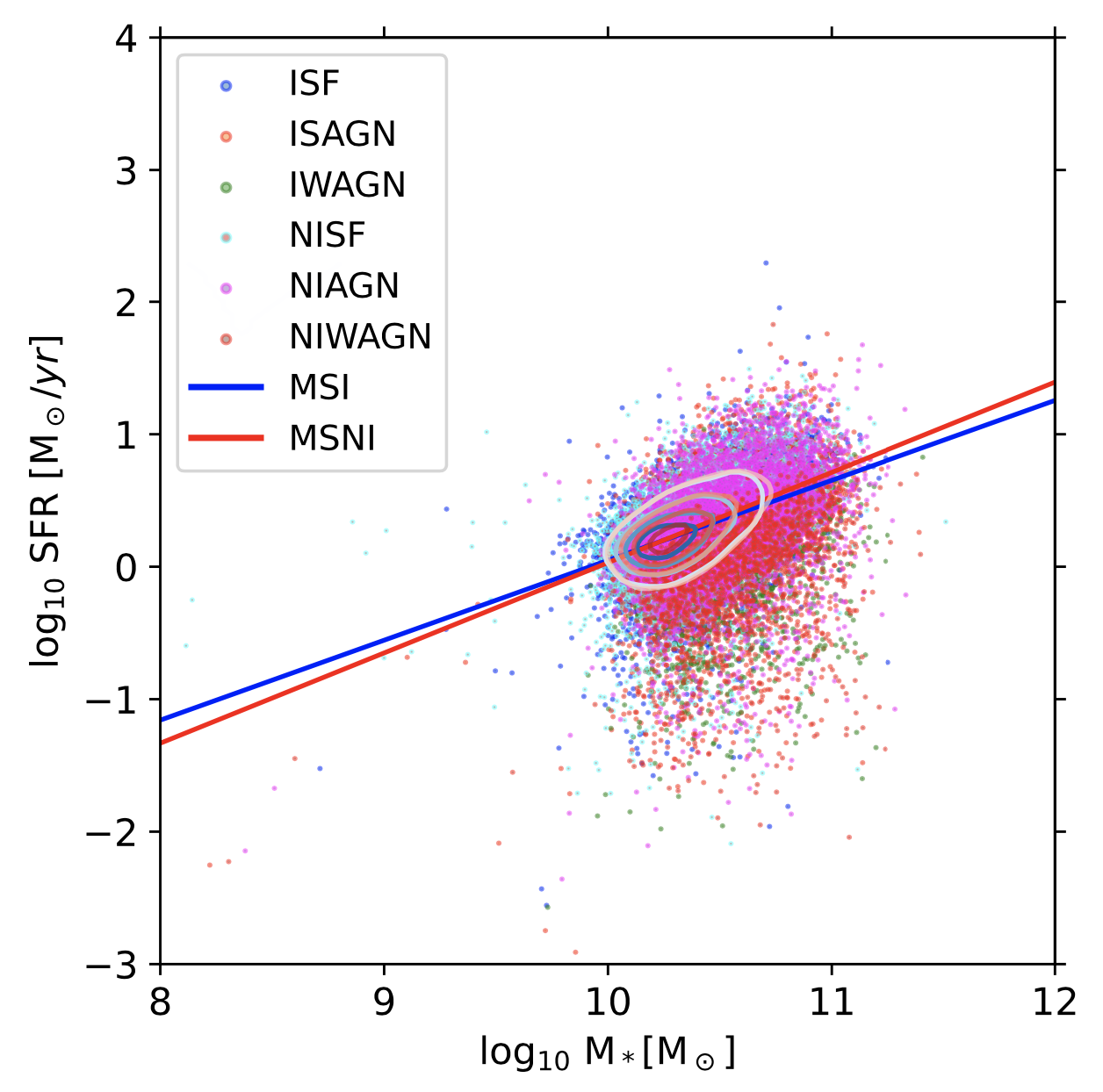}
			\label{MSM}
		}
		\vspace{-0.2cm}
\caption{Star formation main sequence (MS) for ageing galaxies: isolated 
sample (top left panel), non-isolated sample (top right panel), isolated and 
non-isolated on the same plot (bottom panel). In plots, MSI and MSNI 
denote the equations of the main sequence for isolated and non-isolated 
galaxies, respectively.}
		\label{MS}
	\end{figure}

The equations of the  main sequence for SF galaxies show a tight relationship 
between the $\log(\mbox{SFR})$ and $\log(M_\star)$, and the relation has been 
studied in a number of works \cite{elbaz2007reversal,speagle2014highly,
leslie2015quenching,daddi2007multiwavelength,yuan2010role,rich2011galaxy,
leslie2015quenching,schawinski2007observational,whitaker2012star,
shimizu2015decreased}. The relationship serves as the tracer on how the stars 
are formed in relation to the stellar mass within a galaxy. Aiming at 
studying how the environment affects the SFR relative to the $M_{\star}$ for 
the ageing and quenching galaxies, and how the relations are affected by the 
nuclear activity, we generate the equation of the best-fitted line used to 
understand the behaviour of SFR relative to $M_{\star}$ for all galaxies 
residing in isolated and non-isolated environments. Fig.~\ref{MS} shows the 
distributions of SFR with respect to $M_{\star}$ of the SF, strong AGN, and 
weak AGN galaxies for isolated and non-isolated cases along with the 
corresponding best-fitted MS line of the SF galaxies. The width of the MS of 
$ \sim \pm\, 0.3$ dex (red dashed lines) in the plots of this figure is 
selected based on the dispersion of the observed MS. Table \ref{AM} indicates 
the percentage change between isolated and non-isolated galaxies above MS, 
within MS and below MS for SF (columns 2 and 3), Strong AGN (columns 4 and 5), 
Weak AGN (columns 6 and 7) galaxies, respectively.
	
By performing the regression analysis for our samples, the general equations of 
the best-fitted line for the isolated and non-isolated galaxies are 
respectively given by
\begin{align}
\log_{10}(\mbox{SFR}) & = 0.60\pm 0.01\,\log_{10}(M_\star) - 5.99\pm0.14,
  \label{SAMI}\\[8pt]
\log_{10}(\mbox{SFR}) & = 0.63\pm 0.01\,\log_{10}( M_\star) -6.28\pm 0.14,
  \label{SAMNI}
\end{align}
where the associated errors are standard deviations in the slope and intercept.

\begin{table}[h!]
  \centering
   \caption{Number of galaxies within (MS), above (Above MS), and below 
(Below MS) the star-forming main sequence for the isolated (I) and 
non-isolated (NI) environments for the ageing transition (AGT) volume-limited 
sample.}
 \vspace{8pt}
		\setlength{\tabcolsep}{0.34pc}
		\begin{tabular}{ccccccc}
			\toprule
			\toprule
			\multicolumn{3}{c}{{Star-forming (\%)}}&\multicolumn{2}{c}{{Strong AGN (\%)}}&\multicolumn{2}{c}{{Weak AGN (\%)}}
			\\
			\cmidrule(lr){2-3} \cmidrule(lr){4-5}\cmidrule(lr){6-7} 
			Position & I & NI& I& NI& I & NI\\
			{(1)} &{(2)}&{(3)}&{(4)}&{(5)}&{(6)}&{(7)}\\
			\midrule
			MS & $6525 (78.04)$ & $6242 (78.53)$ & $4709 (67.68)$ & $4754 (67.75)$ & $610 (37.89)$ & $577 (37.06)$  \\[2pt]
			
			Above MS & $958 (11.46)$ & $886 (11.15)$ & $782 (11.24)$ & $ 825 (11.76)$ & $1 (0.06)$ & $1 (0.06)$ \\[2pt]
			
			Below MS & $878 (10.50)$ & $821 (10.33)$ & $1467 (21.08)$ & $1438 (20.49)$ & $999 (62.05)$ & $979 (62.88)$  \\[2pt]
			Total	& $8361 (100)$ & $7949 (100)$ & $6958 (100)$ & $7017 (100)$ & $ 1610 (100)$ & $1557 (100)$ \\
			\bottomrule
		\end{tabular}
		\label{AM}
\end{table}

\begin{figure}[t!]
	\subfigure{
		\includegraphics[width=0.45\linewidth]{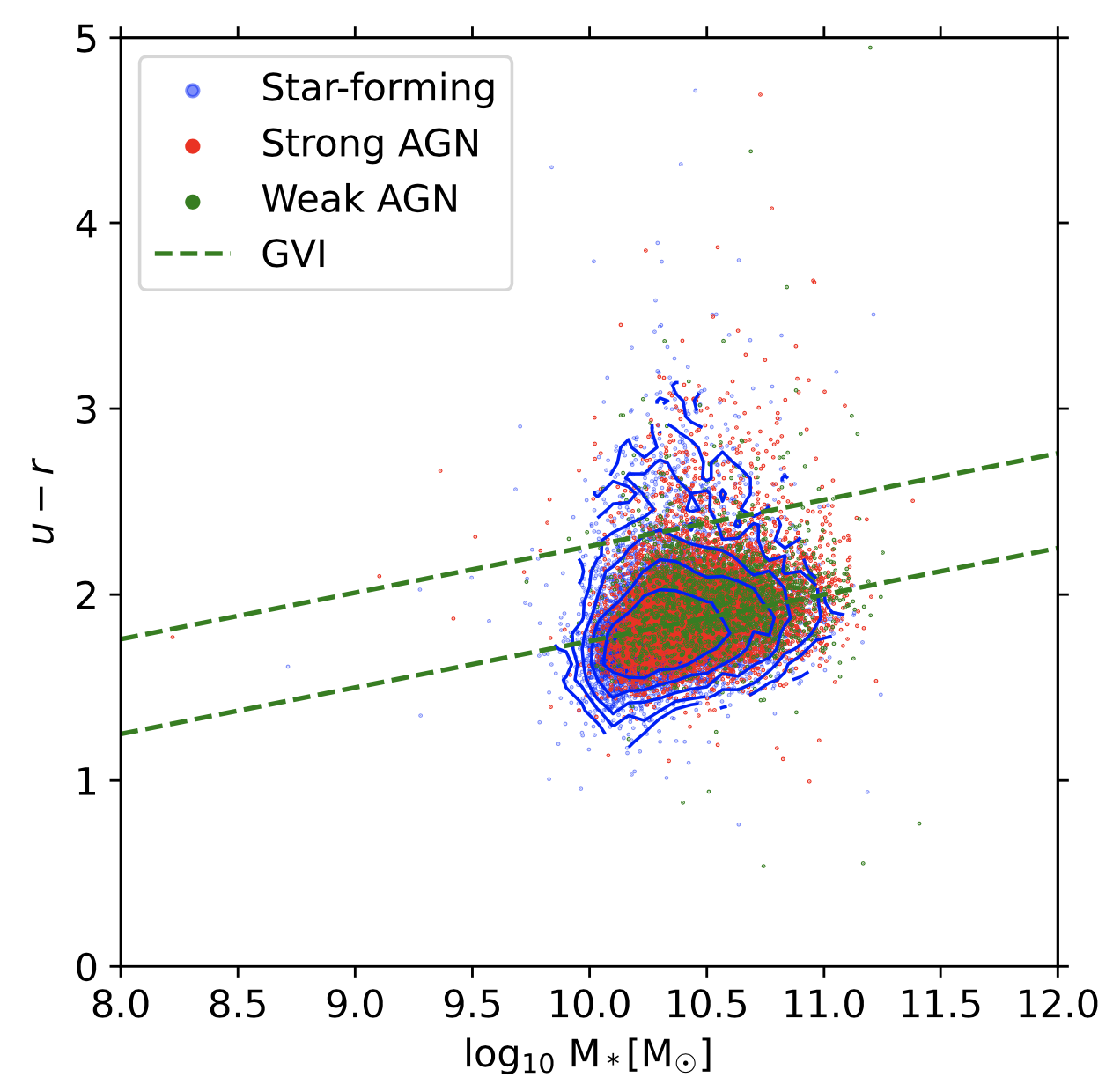}
			\label{CMisolated}
		}
		\hspace{0.1cm}
		\subfigure{
			\includegraphics[width=0.45\linewidth]{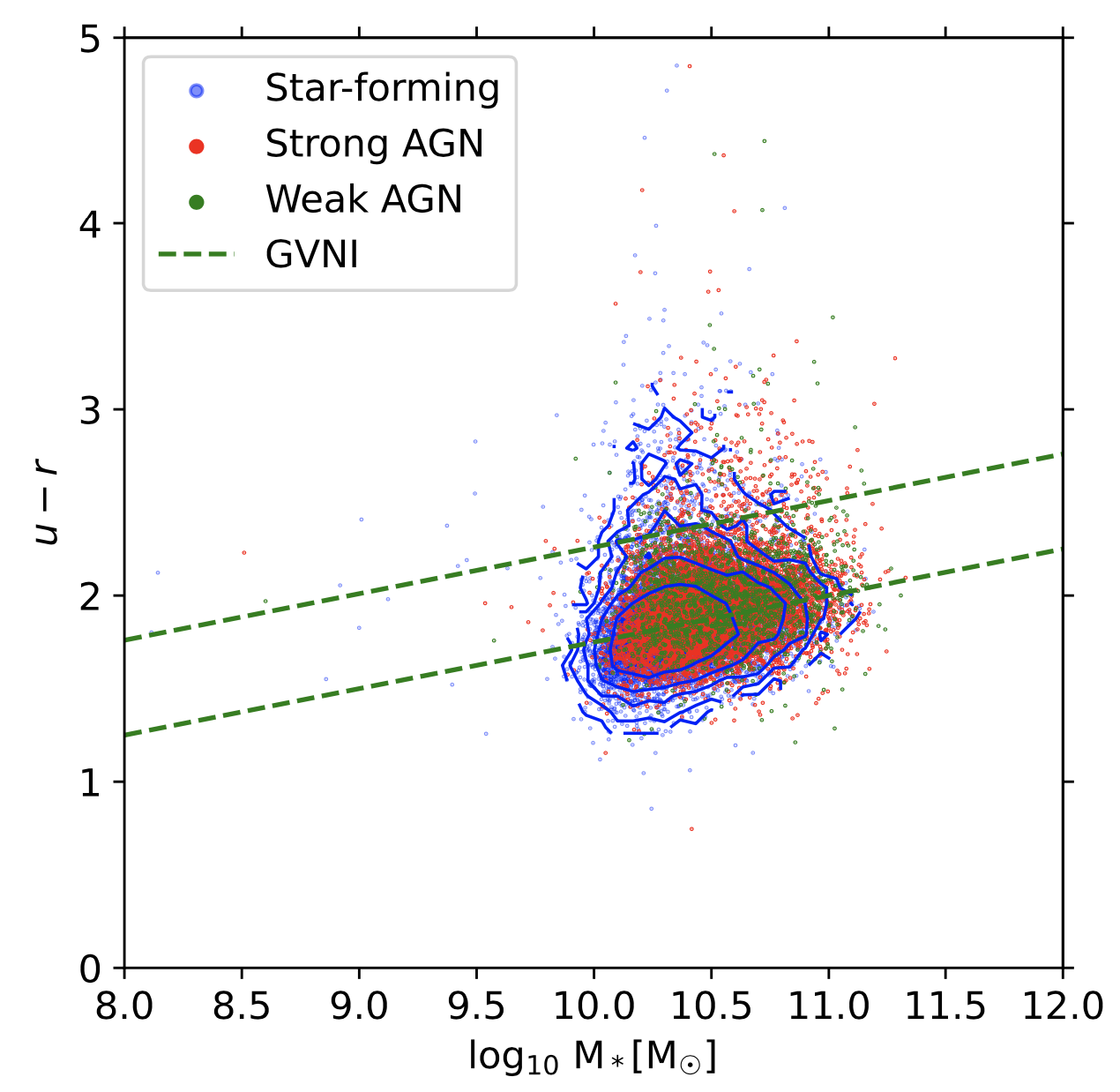}
			\label{CMnonisolated}
		}
		\vspace{-0.2cm}
		\caption{Ageing galaxies' $u-r$ colour against 
			stellar mass diagrams for isolated (left panel) and 
			non-isolated (right panel) samples. Here, GVI 
and GVNI denote the green valleys for isolated and non-isolated galaxies, 
respectively.}
	\label{CM1}
\end{figure}
Galaxies can be categorised further into two groups: those actively forming 
stars, appearing blue, and those lacking significant star formation, appearing 
red. The galaxies initially fall into the blue sub-category and then they 
change gradually to red \cite{gonccalves2012quenching,moustakas2013primus}. 
It is clear to say that evolution from one category to another must involve 
processes that quench their rate of forming new stars from the blue cloud 
passing the intermediate stage (green valley) to the red sequence 
\cite{faber2007galaxy,hickox2014black}. The factors for this transformation 
may be due to internal mechanisms like negative feedback from the AGNs and 
the galaxy environment. To study the effect of the environment on colour, 
Table \ref{AGGV} shows the number of galaxies with respect to the Green 
Valley (GV) for ageing galaxies. The position of galaxies on the colour 
against the stellar mass diagram is shown in Fig~\ref{CM1} for the SF, strong 
AGN, and weak AGN  galaxies respectively. The width of the GV is derived 
following the criteria from Ref.\ \cite{schawinski2014green}, given by equations \eqref{gv1} and  \eqref{gv2} representing the upper and lower lines, respectively as
\begin{align}
	u-r & = -\,0.24 + 0.25 \times M_\star,
		\label{gv1}\\[5pt]
	u-r & = -\,0.75 + 0.25 \times M_\star.
		\label{gv2}
\end{align}
Here $u$ and $r$ magnitudes were derived from the SDSS database with 
extinction corrected. Table \ref{AGGV} indicates the  positioning of galaxies 
with respect to the GV with its percentage for isolated and non-isolated 
galaxies for the ageing sample above GV, within GV and below GV for SF 
(columns 2 and 3), strong AGN (columns 4 and 5), weak AGN (columns 6 and 7) 
galaxies, respectively. 
\begin{table}[h!]
\centering
\caption{Number of galaxies within the green valley (GV), above the green 
valley (Above GV) and below the green valley (Below GV) for isolated (I) and 
non-isolated (NI) galaxies for the ageing sample.}
		\vspace{10pt}
		\setlength{\tabcolsep}{0.4pc}
		\begin{tabular}{ccccccc}
			\toprule
			\toprule
			\multicolumn{1}{c}{} &
			\multicolumn{2}{c}{{Star-forming (\%)}}&\multicolumn{2}{c}{{Strong AGN (\%)}}&\multicolumn{2}{c}{{Weak AGN (\%)}}
			\\
			\cmidrule(lr){2-3} \cmidrule(lr){4-5}\cmidrule(lr){6-7} 
			Position & I & NI& I& NI& I & NI\\
			{(1)} &{(2)}&{(3)}&{(4)}&{(5)}&{(6)}&{(7)}\\
			\midrule
			GV& $3334 (39.88)$ & $3428 (43.12)$ & $3674 (52.80)$ & $3800 (54.15)$ & $920 (57.14)$ & $931 (59.79)$  \\[2pt]
			
			Above GV & $386 (4.62)$ & $424 (5.33)$ & $319 (4.58)$ & $ 340 (4.85)$ & $90 (5.59)$ & $96 (6.17)$  \\[2pt]
			
			Below GV & $4641 (55.51)$ & $4097 (51.54)$ & $2965 (42.61)$ & $2877 (41.00)$ & $600 (37.27)$ & $530 (34.04)$  \\[2pt]
			
	   Total	& $8361 (100)$ & $7949 (100)$ & $6958 (100)$ & $7017 (100)$ & $ 1610 (100)$ & $1557 (100)$ \\
			\bottomrule
		\end{tabular}
		\label{AGGV}
	\end{table}

\subsection{Quenching}
\begin{figure}[t!]
	\subfigure{
		\includegraphics[width=0.46\linewidth]{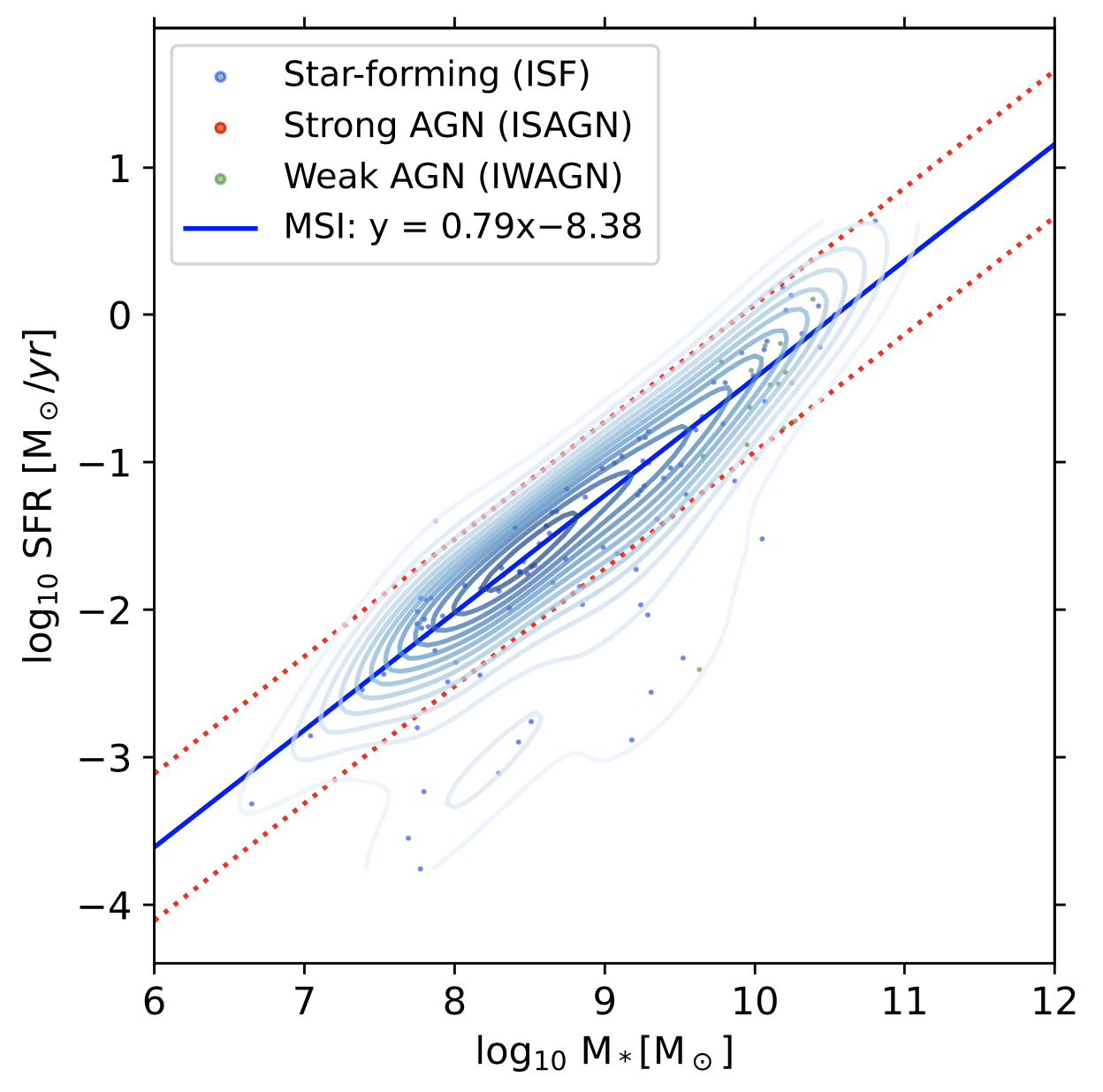}
			\label{MSQI}
		}
		\hspace{0.1cm}
		\subfigure{
			\includegraphics[width=0.46\linewidth]{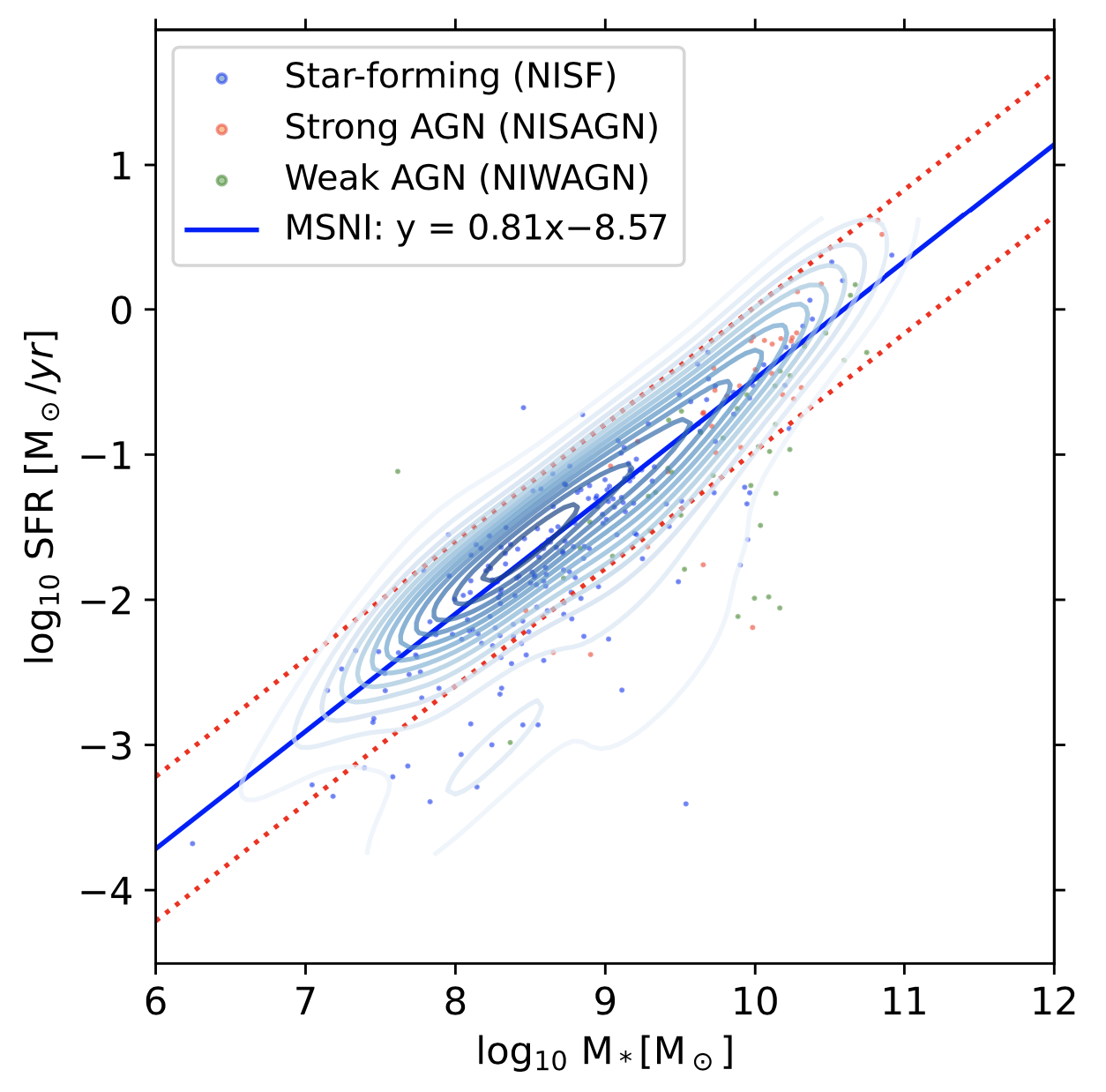}
			\label{MSQNI}
		}
		\hspace{0.1cm}
		\subfigure{
			\includegraphics[width=0.46\linewidth]{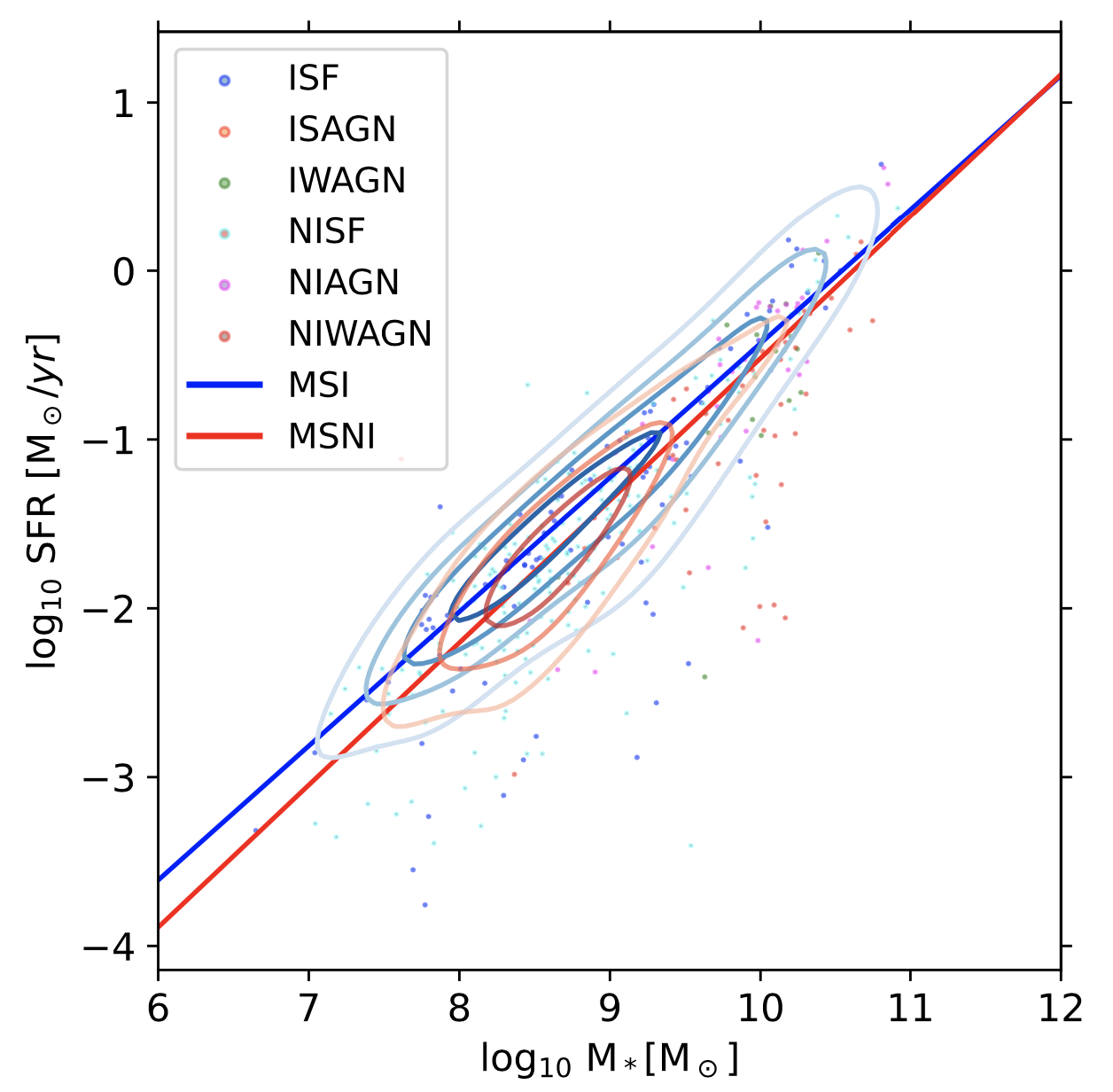}
		}
		\vspace{-0.2cm}
		\caption{Star formation main sequence (MS) for quenching galaxies: isolated sample 
			(top left panel), non-isolated sample (top right panel), isolated and non-isolated on 
			the same plot (bottom panel). In plots, MSI and
MSNI denote the equations of the main sequence for isolated and non-isolated 
galaxies, respectively.}
		\label{MSQ}
	\end{figure}

As we have done in the previous subsection above, we do the same for the 
quenching galaxies. Fig.~\ref{MSQ}, shows the distributions of SFR with 
respect to $M_{\star}$ of the SF, strong AGN, and weak AGN galaxies for the 
isolated and non-isolated cases along with the corresponding best-fitted MS 
line of the SF for quenching galaxies. 
By performing regression analysis for our samples, the general equations of 
the best-fitted line for the isolated and non-isolated galaxies are 
respectively given by 
\begin{align}
    \log_{10}(\mbox{SFR}) & = 0.79\pm 0.06\,\log_{10}(M_\star) -8.38\pm0.52,
        \label{SQMI}\\[8pt]
    \log_{10}(\mbox{SFR}) & = 0.81\pm 0.04\,\log_{10}( M_\star) - 8.57\pm 0.34,
        \label{SQMNI}
\end{align}
where the associated errors are standard deviations in the slope and 
intercept. Table \ref{QMS} indicates the number of galaxies with respect to 
the MS along with the percentage for the isolated and non-isolated 
galaxies above MS, within MS and below MS for SF (columns 2 and 3), strong 
AGN (columns 4 and 5), weak AGN (columns 6 and 7) galaxies, respectively.
	
\begin{figure}[t!]
		\subfigure{
			\includegraphics[width=0.45\linewidth]{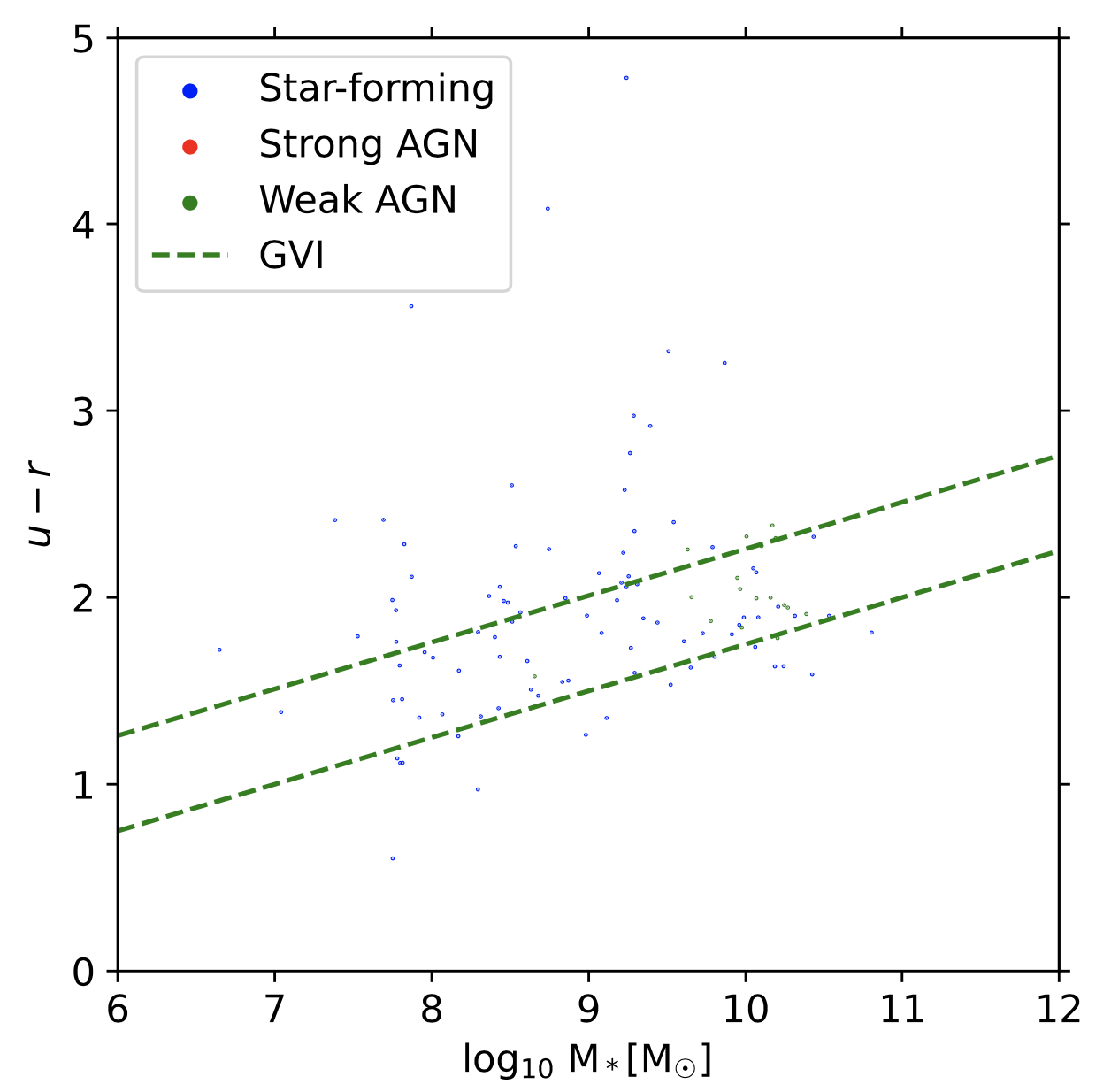}
			\label{CM2isolated}
		}
		\hspace{0.1cm}
		\subfigure{
			\includegraphics[width=0.45\linewidth]{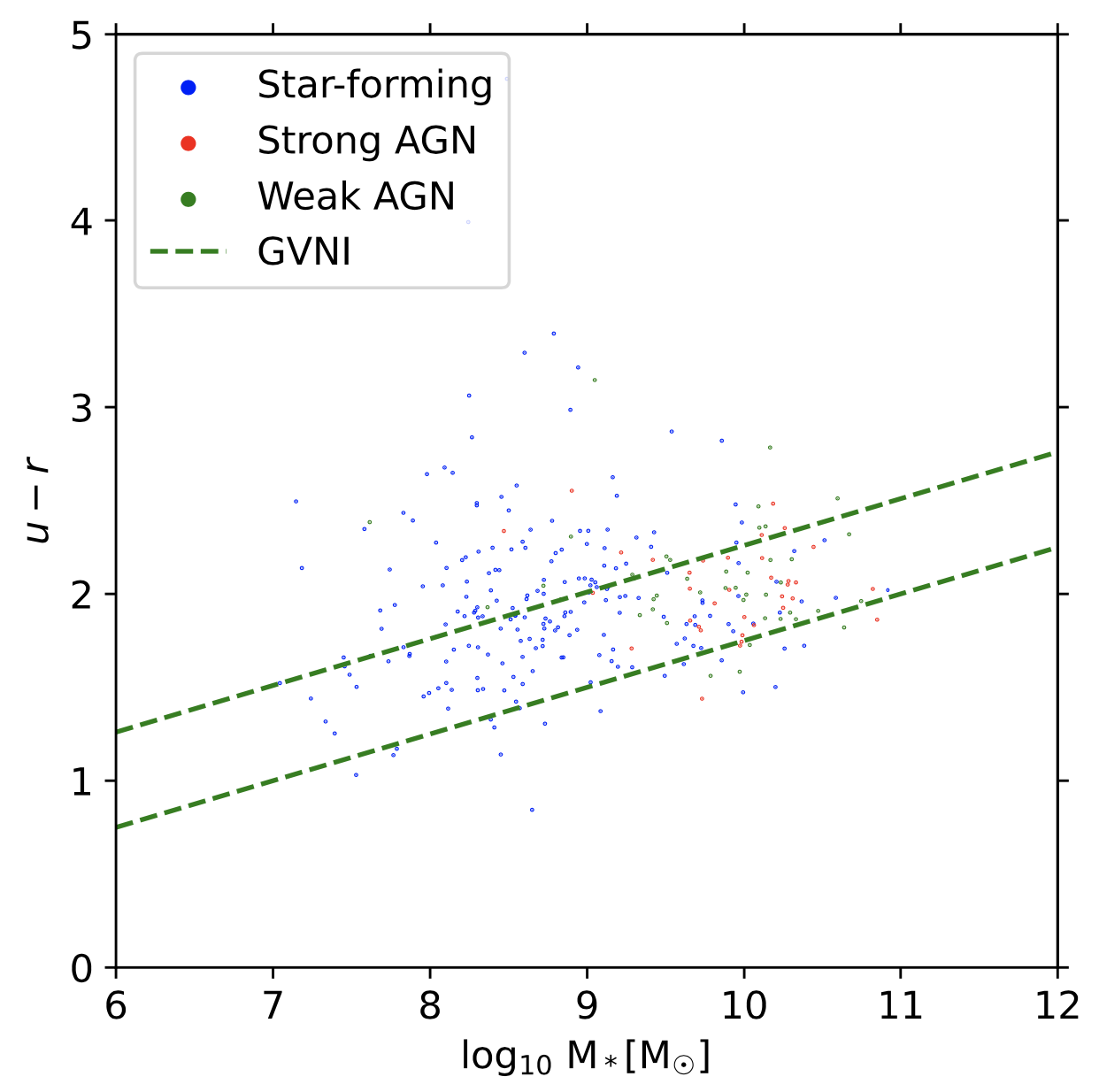}
			\label{CM2nonisolated}
		}
		\vspace{-0.2cm}
		\caption{Quenching galaxies' $u-r$ colour against 
			stellar mass diagrams for the isolated (left panel) 
                        and non-isolated (right panel) samples. 
Here, GVI and GVNI denote the green valleys for isolated and 
non-isolated galaxies, respectively.}
		\label{CM2}
	\end{figure}

\begin{table}[h!]
      \centering
	\caption{Number of galaxies within (MS), above (Above MS), and 
		 below (Below MS) the star-forming main sequence for isolated 
		(I) and non-isolated (NI) environments for the quenching 
                transition (QGT) volume-limited sample.}
		\vspace{10pt}
		\setlength{\tabcolsep}{0.34pc}
		\begin{tabular}{ccccccc}
			\toprule
			\toprule
			\multicolumn{1}{c}{} &
			\multicolumn{2}{c}{{Star-forming (\%)}}&\multicolumn{2}{c}{{Strong AGN (\%)}}&\multicolumn{2}{c}{{Weak AGN (\%)}}
			\\
			\cmidrule(lr){2-3} \cmidrule(lr){4-5}\cmidrule(lr){6-7} 
			Position & I & NI& I& NI& I & NI\\
			{(1)} &{(2)}&{(3)}&{(4)}&{(5)}&{(6)}&{(7)}\\
			\midrule
			MS & $78 (82.98)$ & $158 (76.33)$ & $17 (77.27)$ & $25 (69.44)$ & $16 (94.12\%)$ & $32 (74.42)$  \\[2pt]

			Above MS & $4 (4.26)$ & $23 (11.11)$ & $2 (9.09)$ & $ 7 (19.44)$ & $0 (0)$ & $1 (2.33)$  \\[2pt]

			Below MS & $12 (12.77)$ & $26 (12.56)$ & $3 (13.64)$ & $4 (11.11)$ & $1 (5.88\%)$ & $10 (23.26)$  \\[2pt]
			
			Total	& $94 (100)$ & $207 (100)$ & $22 (100)$ & $36 (100)$ & $ 17 (100)$ & $43(100)$ \\
			\bottomrule
	\end{tabular}
	  \label{QMS}
\end{table}
The position of galaxies on the colour against the stellar mass diagram is 
shown in Fig.~\ref{CM2} for the SF, strong AGN, and weak AGN galaxies 
respectively. A already stated the width of the GV is given by Eqs.~\eqref{gv1} and 
\eqref{gv2}, representing the upper and lower lines, respectively. Table \ref{QGGV} 
indicates the  positioning of galaxies with 
respect to the GV with its percentage for isolated and non-isolated galaxies 
for the quenching sample above GV, within GV and below GV for SF (columns 2 
and 3), strong AGN (columns 4 and 5), weak AGN (columns 6 and 7) galaxies, 
respectively. 
	\begin{table}[h!]
		\centering
		\caption{Number of galaxies within the green valley (GV), 
                         above the green valley (Above GV) and below the green 
                         valley (Below GV) for isolated (I) and non-isolated 
                         (NI) galaxies in the quenching sample.}
		\vspace{10pt}
		\setlength{\tabcolsep}{0.34pc}
		\begin{tabular}{ccccccc}
			\toprule
			\toprule
			\multicolumn{1}{c}{} &
			\multicolumn{2}{c}{{Star-forming (\%)}}&\multicolumn{2}{c}{{Strong AGN (\%)}}&\multicolumn{2}{c}{{Weak AGN (\%)}}
			\\
			\cmidrule(lr){2-3} \cmidrule(lr){4-5}\cmidrule(lr){6-7} 
			Position &I & NI& I& NI& I & NI\\
			{(1)} &{(2)}&{(3)}&{(4)}&{(5)}&{(6)}&{(7)}\\
			\midrule
			GV & $42 (44.68)$ & $95 (45.89)$ & $13 (59.09)$ & $24 (66.67)$ & $12 (70.59\%)$ & $25 (58.14)$  \\[2pt]
			
			Above GV & $36 (38.30)$ & $95 (45.89)$ & $2 (9.09)$ & $ 8 (22.22)$ & $4 (23.53\%)$ & $14 (32.56)$  \\[2pt]
			
			Below GV & $16 (17.02)$ & $17 (8.21)$ & $7 (31.82)$ & $4 (11.11)$ & $1 (5.88\%)$ & $4 (9.30)$  \\[2pt]
			
			Total	& $94 (100)$ & $207 (100)$ & $22 (100)$ & $36 (100)$ & $ 17 (100)$ & $43(100)$ \\ 
			\bottomrule
		\end{tabular}
		\label{QGGV}
	\end{table}
Table \ref{CPMS} indicates the Chi-square P-values between isolated and 
non-isolated galaxies above MS, within MS and below MS for SF (column 2), 
strong AGN (column 3), weak AGN (column 4) ageing galaxies, respectively and 
SF (column 5), strong AGN (column 6), weak AGN (column 7) quenching galaxies, 
respectively. Table \ref{CPGV} indicates the Chi-square P-values between 
isolated and non-isolated galaxies above GV, within GV and below GV for SF 
(column 2), strong AGN (column 3), weak AGN (column 4) ageing galaxies,
 respectively and SF (column 5), strong AGN (column 6), weak AGN (column 7) 
quenching galaxies, respectively. These tables are used to trace how the 
positioning of galaxies with respect to MS and GV are affected by changing the 
galaxies' environment and the influence of nuclear activity on the relation.
	\begin{table}[!h]
		\centering
		\caption{Chi-square P-values for ageing and quenching 
transitions between isolated and non-isolated galaxies within MS, above MS, 
and below MS. }
		\vspace{8pt}
		\setlength{\tabcolsep}{0.5pc}
		\scalebox{1}{
			\begin{tabular}{cccc|ccc}
				\toprule
				\toprule
				& \multicolumn{3}{c}{Ageing  } & \multicolumn{3}{c}{Quenching } \\
				\cmidrule(lr){2-4} \cmidrule(lr){5-7}
				Position & Star-forming & Strong AGN & Weak AGN & Star-forming & Strong AGN & Weak AGN \\
				{(1)} & {(2)} & {(3)} & {(4)} & {(5)} & {(6)} & {(7)}  \\
				\midrule
				MS &  \num{0.018} & \num{0.350} & \num{1} & \num{0.087}& \num{0.495} & \num{0.633} \\[2pt]
				Above MS & \num{0.001} & \num{0.942}& \num{0.656}& \num{0.251}& \num{0.730}& \num{0.072} \\[2pt]	
				Below MS & \num{0.014}& \num{0.401} & \num{0.657}& \num{1}& \num{1} & \num{0.006} \\[2pt]
				\bottomrule
			\end{tabular}
		}
		\label{CPMS}
	\end{table}
	\begin{table}[!h]
		\centering
		\caption{Chi-square P-values for ageing and quenching 
transitions between isolated and non-isolated galaxies within GV, above GV, 
and below GV.}
		\vspace{10pt}
		\setlength{\tabcolsep}{0.5pc}
		\scalebox{1}{
			\begin{tabular}{cccc|ccc}
				\toprule
				\toprule
				& \multicolumn{3}{c}{Ageing } & \multicolumn{3}{c}{Quenching } \\
				\cmidrule(lr){2-4} \cmidrule(lr){5-7}
				Position & Star-forming & Strong AGN & Weak AGN & Star-forming & Strong AGN & Weak AGN \\
				{(1)} & {(2)} & {(3)} & {(4)} & {(5)} & {(6)} & {(7)}  \\
				\midrule
				GV & \num{0} & \num{0.113} & \num{0.139} & \num{0.943} & \num{0.763} & \num{0.928} \\[2pt]
				Above GV & \num{0.038} & \num{0.492} & \num{0.540} & \num{0.269} & \num{0.354} & \num{0.830}\\[2pt]    
				Below GV & \num{0} & \num{0.055} & \num{0.063} & \num{0.269} & \num{0.108} & \num{0.310} \\[2pt]
				\bottomrule
			\end{tabular}
		}
		\label{CPGV}
	\end{table}

\section{Discussion}
\label{secV}
From Fig.~\ref{WHAN}, it is observed that the number of star-forming galaxies 
decreases by $\sim 12\%$ between isolated and non-isolated environments, while 
the retired galaxies increase by $\sim 14\%$, the number of strong AGN decreases 
by $\sim 3\%$ and the weak AGN increase by $\sim 1\%$. This indicates that the 
star-forming and retired galaxies are more significantly affected by the 
environment than the AGNs, hence environmental dependence is influenced by the 
presence of AGN sources. From Fig.~\ref{AD} it is observed that the ageing 
galaxies decrease by $\sim 14\%$, while the retired galaxies increase by 
$\sim 14 \%$. The undetermined galaxies decrease by $\sim 0.2\%$ while quenched 
galaxies increase by $\sim 0.3\%$ between isolated and non-isolated environments. 
This proves the scenario of Fig.~\ref{WHAN}, where the ageing (which mostly 
contains star-forming galaxies, see Ref.~\cite{corcho2023ageinga}) and retired 
galaxies are strongly affected by the environment than undetermined and 
quenched galaxies. This trend shows that quenched and retired galaxies are 
found in non-isolated environments rather than isolated environments while 
ageing is primarily found in isolated rather than non-isolated environments. 
The study agrees with Ref.\ \cite{corcho2023ageingb} using the density approach 
that quenched galaxies are primarily found in dense while ageing galaxies are 
found in less dense environments.
	
From Fig.~\ref{MS} and Eqs.~\eqref{SAMI} and \eqref{SAMNI}, the difference of 
$\sim 0.03$ in slope and $\sim 0.30$ in intercept are observed between isolated and 
non-isolated environment for ageing galaxies. The differences in slope and 
intercept are greater than the errors in slopes ($0.01$) and intercept 
($0.14$). Furthermore, these differences produce P-values of $\num{5.190e-05}$, and 
$\num{6.408e-05}$ for the slope and intercept, respectively which are both 
far less than the standard P-value in statistics ($0.05$). These two facts 
indicate that there is a significant difference in slope and intercept between 
isolated and non-isolated environments for transition ageing galaxies.  
The steeper slope ($0.63$) and lower intercept ($-\,6.28$) of 
non-isolated galaxies when compared to isolated galaxies with slope of 
$0.60$ and intercept of $-\,5.99$ in Fig.~\ref{MS}, indicates that 
environmental factors affect the decrease of SFR in relation to stellar mass 
in ageing galaxies. This increases the transition of blue cloud to red 
sequence shaping the observed colour-mass bi-modality in Fig.~\ref{CM1}, which 
is more rapid in the non-isolated than the isolated environments. The observed 
difference is evident in Table \ref{AGGV} where the non-isolated galaxy 
fraction is higher than isolated. Combining the observations in both 
figures it is evident that the environment is responsible for the decrease of 
SFR and the galaxy transition on the colour-stellar mass diagram.
	
From Fig.~\ref{MSQ} and Eqs.~\eqref{SQMI} and \eqref{SQMNI}, the difference of 
$0.02$ in slope and $0.19$ in intercept are observed between isolated and 
non-isolated environment for quenching galaxies. The difference in slope and 
intercept are less than the errors in slopes ($0.06$, $0.04$) and intercepts 
($0.52$, $0.34$). Furthermore, these differences produce P-values of $0.50$, 
and $0.36$ for the slope and intercept, respectively which are both greater 
than the standard P-value in statistics ($0.05$). These two facts indicate 
that there is no significant difference in slope and intercept between 
isolated and non-isolated environments for quenching galaxies.  
	
From Tables \ref{AM}, \ref{AGGV}, \ref{QMS}, and \ref{QGGV} it is observed 
that the number of SF, strong AGN, and weak AGN with respect to the MS have 
different fractions for the isolated and non-isolated environments in both 
ageing and quenching galaxies. In this observation the study agrees with the 
findings by Ref.\ \cite{oemler2017star}, that galaxies classified as MS, 
quiescent and passive exhibit different fractions across different 
environments. On the other hand, the study agrees with 
Ref.\ \cite{corcho2023ageingb}, that ageing and quenching galaxies are 
found in different environments.
	
From column (2)  of Table \ref{CPMS}, it is observed that the chi-square 
($\chi^{2}$) P-value for the difference in numbers of ageing galaxies between 
isolated and non-isolated within MS, above MS, and below MS is smaller 
($\sim \num{0.011}$ on average) than the threshold ($\num{0.05}$) while from 
column (3) and (4) of Table \ref{CPMS}, it is observed that the P-values are 
greater ($\sim \num{0.564}, \sim \num{0.771}$ on average) than the threshold 
($\num{0.05}$). On the other hand from columns (5), (6), (7) of Table 
\ref{CPMS}, it is observed that the P-values are greater ($\sim \num{0.446}, 
\sim \num{0.742},\sim \num{0.237}$ on average) than the threshold. Based on 
these observations, the positioning of SF ageing galaxies along the MS is 
affected by the environment, while the AGNs are not affected by the 
environment. The positioning of quenching galaxies is not influenced by the
environment and this trend does not depend on the nature of nuclear activities 
of a particular galaxy.
	
From column (2) of Table \ref{CPGV}, it is observed that the chi-square 
($\chi^{2}$) P-value for the difference in numbers of ageing galaxies between 
isolated and non-isolated within MS, above MS, and below MS is smaller 
($\sim \num{0.013}$ on average) than the threshold ($0.05$) while from columns
(3) and (4) of Table \ref{CPGV}, it is observed that the P-values are greater 
($\sim \num{0.220}, \sim \num{0.247}$ on average) than the threshold. On the 
other hand from columns (5), (6), (7) of Table \ref{CPGV}, it is observed that 
the P-values are greater ($\sim \num{0.494}, \sim \num{0.408}, 
\sim \num{0.689}$ on average) than the threshold. All these observations 
imply that the positioning of SF ageing galaxies with respect to the GV is 
affected by the environment, while the AGNs are not affected by the environment. 
The positioning of quenching galaxies with respect to the GV is not influenced 
by the environment and again this trend is not affected by the nature of 
nuclear activities of a particular galaxy.

It is widely believed that environmental influence is a long-term 
process which takes a few billion years. Even mergers effects do not happen 
soon after galaxies' merging event, the median delay is $\sim1.5$ Gyr where the 
time scale varies over a long range \cite{peng2015strangulation}. Due to the 
time scale factor, ageing is expected to be favoured by the environment rather 
than quenching. For example, the study in Ref.\ \cite{mao2022revealing} 
revealed 
that the quenching to depend on mass and not environment using the 17 galaxy 
cluster candidates from the Cosmic Evolution Survey (COSMOS) field. Using the 
Sims Implementing Metals and Blackholes in Astrophysics (SIMBA)
cosmological simulation to find the relationship between galaxy merging 
(related to environment) and quenching, Ref.\ \cite{rodriguez2019mergers} 
observed that the distribution of time delay between merging and quenching 
events does not suggest any physical relationship with fast or slow quenching. 
SIMBA also predicts that a major merger will trigger starbursts but it is not 
related to quenching in fast or slow mode. Ref.\ \cite{tan2022resolved}, 
revealed that the effect of the environment on the galaxy quenching is not 
completely separated from the effect of mass on galaxy quenching. This implies 
that mass is the leading parameter in the quenching process 
\cite{ge2020conditions,reeves2021gogreen,corcho2021galaxy,hasan2023evolving,kipper2021role,einasto2022death}.

All necessary procedures have been considered in the analysis 
remembering the existence of observational bias, selection effect and 
methodological limitations. These include the bias from SDSS fibre collision 
and incomplete sky coverage which exclude low luminosity galaxies in the 
dense region explained in Section~\ref{sdss}. The creation of the volume 
limited sample assured the uniformity in luminosity which is the key aspect to 
minimize bias explained in Section \ref{vl}. The use of the friends-of-friends 
method is subject to uncertainty especially on the selection of the 
linking length which may over-link or under-link the galaxies affecting the 
analysis as explained in Section \ref{env}. Lastly, the uncertainty due to 
statistical analysis in the measurements of galaxy properties also is another 
important area to address as explained in Section \ref{pro}. All these have 
been taken into account to minimize errors which may otherwise significantly 
influence the results.
	
\section{Summary and Conclusion}
	\label{secVI}
This study aimed at investigating if the environment is among the factors 
affecting the slow decrease (ageing), and fast decrease (quenching) of SFR in 
galaxies and how the relationship is influenced by the nuclear activity of a 
particular galaxy at the transition stage. We used the friends-of-friends 
method (defined in Section \ref{secIII}) to assign the volume-limited sample 
from SDSS DR 12 galaxies into isolated and non-isolated environments. From a 
flux-limited sample of SDSS DR12, we constructed the volume-limited sample of 
136274 galaxies within $ -22.5\leq M_r \leq -20.5$ (mag), containing 58032 
($\sim 43\%$) isolated and 78233 ($\sim 57\%$) non-isolated galaxies. The galaxies 
were classified according to their nature of nuclear activity employing WHAN 
diagnostic diagrams (shown by Fig.~\ref{WHAN}) into star-forming, strong AGN, 
weak AGN, and retired galaxies aiming to investigate the influence of nuclear 
activity on the environmental dependence of galaxies.  Ageing diagrams (shown 
by Fig.~\ref{AD}) were used to classify the galaxies into ageing, quenched, 
undermined, and retired galaxies. Since we aim to study the ageing and 
quenching galaxies in the transition stage,  we used the $0.5 < (g - r) < 0.7$, 
criteria from Ref.\ \cite{corcho2023ageinga}, to obtain a total number of  
$33452$ ageing, galaxies in transition stage (AGT) where $16929$ are isolated 
and $16523$ are non-isolated. Furthermore, a total number of $419$ quenching 
(QGT) was obtained, where $133$ are isolated and $286$ are non-isolated. The 
AGT and QGT were used throughout the study to investigate the star formation 
main sequence and the colour stellar mass diagram for ageing galaxies 
(shown by Figs.~\ref{MS} and~\ref{CM1}) and quenching galaxies (shown by 
Figs.~\ref{MSQ} and~\ref{CM2}). The following significant findings are 
revealed via this study:
\begin{itemize}
\item Ageing galaxies are mostly found in isolated rather than the non-isolated 
environment while quenching galaxies are found in non-isolated rather than 
the isolated environment.
\item  The significant change in slope and intercept of the equation of star 
formation main sequence by $0.03$ dex, $0.30$ dex respectively between 
isolated and non-isolated environments for ageing galaxies was observed 
indicating that the slope and intercept of ageing star formation main sequence 
are influenced by the environment.
\item  The insignificant change in slope and intercept of the equation of star 
formation main sequence by $0.02$ dex, $0.12$ dex respectively between 
isolated and non-isolated environments for quenching galaxies was observed 
indicating that the slope and intercept of quenching star formation main 
sequence are not influenced by the environment.
\item A significant change in the number of ageing SF galaxies above, within 
and below the main sequence, similarly above, within and below the green valley 
between isolated and non-isolated environments was observed indicating that the 
positioning of ageing star forming galaxies with respect to the main sequence 
and the green valley is influenced by the environment. On the other hand 
insignificant change in the number of ageing strong and weak AGN was observed 
indicating that the positioning of ageing AGN are not influenced by the
environment. This implies that ageing depends on the environment and this 
relationship is influenced the nuclear activity.
\item An insignificant change in the number of quenching SF, strong AGN, and 
weak AGN was observed above, within and below the main sequence, similarly  
above, within and below the green valley between isolated and non-isolated 
environments was observed indicating that the positioning of quenching star 
forming galaxies with respect to the main sequence and the green valley are 
not influenced by the environment and the nuclear activity have negligible 
influence on the independence of quenching on the environment.
\end{itemize}
Altogether, the results of this study help us to understand galaxy 
evolution by revealing how environmental factors and nuclear activity affect 
the galaxy's transitions. The significant environmental dependence of ageing 
proves the results from theoretical predictions on the role of environmentally 
driven processes such as the ram pressure stripping able to remove the gas 
reservoir, leading to strangulation or starvation (the suppression of gas 
infall or galaxy interactions) in the star formation rate of galaxies. The 
independence of quenching on the environment proves that internally triggered 
mechanisms such as negative feedback from AGNs, supernovae winds or 
the stabilisation of the gas against fragmentation lead to the rule on the 
quenching
process. The difference in slopes and intercepts between ageing and quenching 
proves that the process of galaxy evolution is not uniform but a multi-process 
indicating the combination of both environmental and internal factors. The 
results challenge the oversimplified notion and highlight the interplay between 
environment, internal galaxy dynamics, and nuclear activity. 

Despite the strength of using the volume limited sample and the 
effectiveness of using the WHAN diagnostic diagram in galaxy classification to 
minimize the selection bias these obviously may exclude some populations such 
as faint, massive and composite galaxies. Furthermore, comparing two 
populations of isolated and non-isolated galaxies may exclude the intermediate 
density environment galaxies which may not significantly change the 
implication of existing results but provide additional insights.

In the future, we will apply similar analysis on the galaxies of high 
redshift surveys example 
the James Webb Space Telescope (JWST) Ref.\ \cite{mcelwain2023james} to again 
characterize the influence of the 
environment on ageing and quenching of galaxies. Further, we plan to use 
integral field spectroscopy (IFS) surveys example Mapping Nearby 
Galaxies survey at APO (MaNGA) Ref.\ \cite{drory2015manga} data to investigate the 
role of morphology-environment relation on galaxy transitions.

\section*{Acknowledgements} PP acknowledges support from The Government of 
Tanzania through the India Embassy, Mbeya University of Science and Technology 
(MUST) for Funding and SDSS for providing data. UDG is thankful to the 
Inter-University Centre for Astronomy and Astrophysics (IUCAA), Pune, India 
for the Visiting Associateship of the institute.  Funding for SDSS-III has been provided by 
the Alfred P. Sloan Foundation, the Participating Institutions, the National Science Foundation, 
and the U.S. Department of Energy Office of Science. 
The SDSS-III website is http://www.sdss3.org/. SDSS-III is managed by the 
Astrophysical Research Consortium for the Participating Institutions of the 
SDSS-III Collaboration including the University of Arizona, the Brazilian Participation Group, 
Brookhaven National Laboratory, Carnegie Mellon University, the University of Florida, 
the French Participation Group, the German Participation Group, Harvard University, 
the Instituto de Astrofisica de Canarias, the Michigan State/Notre Dame/JINA Participation Group, 
Johns Hop- kins University, Lawrence Berkeley National Laboratory, Max Planck Institute for Astrophysics, 
Max Planck Institute for Extraterrestrial Physics, New Mexico State University, 
New York University, Ohio State University, Pennsylvania State University, 
University of Portsmouth, Princeton University, the Spanish Participation Group, University of Tokyo, 
the University of Utah, Vanderbilt University, the University of Virginia, the University of Washington, and Yale University.


\begin{thebibliography}{99}
	\bibitem[Kauffman et al.(2004)]{kauffmann2004environmental}
	G. Kauffmann, S. White, \emph{The Environmental Dependence of the Relations between Stellar Mass, Structure, Star Formation and Nuclear Activity in Galaxies}, \href{https://doi.org/10.1111/j.1365-2966.2004.08117.x}{MNRAS \textbf{353}, 713 (2004)} [\href{https://arxiv.org/abs/astro-ph/0402030}{arXiv:astro-ph/0402030}].
	
	\bibitem[Tremonti et al.(2004)]{tremonti2004origin}
	C. Tremonti, T. Heckman, \emph{The Origin of the Mass--Metallicity Relation: Insights from 53,000 Star-Forming Galaxies in the SDSS}, \href{https://doi.org/10.1086/423264}{ApJ \textbf{613},  898 (2004)}   [\href{https://arxiv.org/abs/astro-ph/0405537}{arXiv:astro-ph/0405537}].
	
	\bibitem[Peng et al.(2010)] {peng2010mass}
	Y. Peng, S. Lilly, K. Kovač, M. Bolzonella, L. Pozzetti, et al., \emph{Mass and Environment as Drivers of Galaxy Evolution in SDSS and zCOSMOS and the Origin of the Schechter Function}, \href{https://doi.org/10.1088/0004-637X/721/1/193}{ApJ \textbf{721}, 193 (2010)} [\href{https://arxiv.org/abs/1003.4747}{arXiv:1003.4747}].
	
	\bibitem[Brinchmann et al.(2004)]{brinchmann2004physical}
	J. Brinchmann, S. Charlot, \emph{The physical properties of star forming galaxies in the low redshift universe}, \href{https://doi.org/10.1111/j.1365-2966.2004.07881.x}{MNRAS \textbf{351}, 1151 (2004)} [\href{https://arxiv.org/abs/astro-ph/0311060}{arXiv:astro-ph/0311060}].
	
	\bibitem{gonccalves2012quenching}	
	T. Gonçalves, D. Martin, K. Menéndez-Delmestre, T. Wyder, et al., \emph{Quenching star formation at intermediate redshifts: downsizing of the mass flux density in the green valley}, \href{https://doi.org/ 10.1088/0004-637X/759/1/67}{ApJ \textbf{759}, 67 (2012)} [\href{https://arxiv.org/abs/1209.4084}{arXiv:1209.4084}].
	
	\bibitem{moustakas2013primus}
	J. Moustakas, A. Coil, J. Aird, M. Blanton, R. Cool, et al., \emph{Constraints on Star Formation Quenching and Galaxy Merging, and the Evolution of the Stellar Mass Function From $z=0-1$}, \href{https://doi.org/10.1088/0004-637X/767/1/50}{ApJ \textbf{767}, 50 (2013)} [\href{https://arxiv.org/abs/1301.1688}{arXiv:1301.1688}].
	
	\bibitem[Elbaz et al.(2007)]{elbaz2007reversal}
	D. Elbaz, E. Daddi, D. Borgne, \emph{The reversal of the star formation-density relation in the distant universe}, \href{https://doi.org/10.1051/0004-6361:20077525}{A\&A \textbf{468}, 33 (2007)}   [\href{https://arxiv.org/abs/astro-ph/0703653}{arXiv:astro-ph/0703653}].
	
	\bibitem[Speagle et al.(2014)]{speagle2014highly}
	J. Speagle, C. Steinhardt, P. Capak, J. Silverman, et al., \emph{A Highly Consistent Framework for the Evolution of the Star-Forming" Main Sequence" from $z\sim 0-6$}, \href{https://doi.org/10.1088/0067-0049/214/2/15}{ApJS \textbf{214}, 15 (2014)} [\href{https://arxiv.org/abs/1405.2041}{arXiv:1405.2041}].
	
	\bibitem{leslie2015quenching}
	S. Leslie, L. Kewley, D. Sanders, N. Lee, \emph{Quenching star formation: insights from the local main sequence}, \href{https://doi.org/10.1093/mnrasl/slv135}{MNRAS \textbf{455}, L82 (2015)} [\href{https://arxiv.org/abs/1509.03632}{arXiv:1509.03632}].
	
	\bibitem{daddi2007multiwavelength}
	E. Daddi, M. Dickinson, G. Morrison, R. Chary, et al., \emph{Multiwavelength study of massive galaxies at $z\sim 2$. I. Star formation and galaxy growth}, \href{https://doi.org/10.1086/521818}{ApJ \textbf{670}, 156 (2007)}.
	[\href{https://arxiv.org/abs/0705.2831}{arXiv:0705.2831}].
	
	\bibitem{yuan2010role}
	T. Yuan, L. Kewley, D. Sanders, \emph{The Role of Starburst-Active Galactic Nucleus Composites in Luminous Infrared Galaxy Mergers: Insights from the New Optical Classification Scheme}, \href{https://doi.org/10.1088/0004-637X/709/2/884}{ApJ \textbf{709}, 884 (2010)}.
	
	\bibitem{rich2011galaxy}
	J. Rich, L. Kewley, M. Dopita, \emph{Galaxy-Wide Shocks in Late-Merger Stage Luminous Infrared Galaxies}, \href{https://doi.org/10.1088/0004-637X/734/2/87}{ApJ \textbf{782}, 9 (2014)} [\href{https://arxiv.org/abs/1104.1177}{arXiv:1104.1177}].		
	
	\bibitem[Schawinski et al.(2007)]{schawinski2007observational}
	K. Schawinski, D. Thomas, M. Sarzi, C. Maraston, et al., \emph{Observational evidence for AGN feedback in early-type galaxies}, \href{https://doi.org/10.1111/j.1365-2966.2007.12487.x}{MNRAS \textbf{382}, 1415 (2007)} [\href{https://arxiv.org/abs/astro-ph/0709.3015}{arXiv:0709.3015}].
	
	\bibitem{whitaker2012star}
	K. Whitaker, P. Dokkum, G. Brammer, M. Franx, \emph{The star formation mass sequence out to z= 2.5}, \href{https://doi.org/10.1088/2041-8205/754/2/L29}{ApJL \textbf{754}, L29 (2015)} [\href{https://arxiv.org/abs/1205.0547}{arXiv:1205.0547}].
	
	\bibitem{shimizu2015decreased}
	T. Shimizu, R. Mushotzky, M. Meléndez, M. Koss, et al., \emph{Decreased specific star formation rates in AGN host galaxies}, \href{https://doi.org/10.1093/mnras/stv1407}{MNRAS \textbf{452}, 1841 (2007)} [\href{https://arxiv.org/abs/1506.07039}{arXiv:1506.07039}].
	
	\bibitem{faber2007galaxy}
	S. Faber, C. Willmer, C. Wolf, D. Koo, B. Weiner, J. Newman, et al., \emph{Galaxy Luminosity Functions to z$\sim$1: DEEP2 vs. COMBO-17 and Implications for Red Galaxy Formation}, \href{https://doi.org/10.1086/519294}{ApJ \textbf{665}, 265 (2007)} [\href{https://arxiv.org/abs/astro-ph/0506044}{arXiv:astro-ph/0506044}].
	
	\bibitem{hickox2014black}
	R. Hickox, J. Mullaney, D. Alexander, C. Chen, F. Civanoet, al., \emph{Black hole variability and the star formation–active galactic nucleus connection: do all star-forming galaxies host an active galactic nucleus}, \href{https://doi.org/10.1088/0004-637X/782/1/9}{ApJ \textbf{782}, 9 (2014)} [\href{https://arxiv.org/abs/1306.3218}{arXiv:1306.3218}].
	
	\bibitem{schawinski2014green}
	K. Schawinski, C. Urry, B. Simmons, L. Fortson, S. Kaviraj, et al., \emph{The green valley is a red herring: Galaxy Zoo reveals two evolutionary pathways towards quenching of star formation in early-and late-type galaxies}, \href{https://doi.org/10.1093/mnras/stu327}{MNRAS \textbf{440}, 889 (2014)}.
	
	\bibitem[Corcho et al.(2023a)]{corcho2023ageinga}	
	P. Corcho-Caballero, Y. Ascasibar, L. Cortese, S. Sanchez, et al., \emph{Ageing and quenching through the Ageing Diagram – II. Physical characterization of galaxies}, \href{ https://doi.org/10.1093/mnras/stad2096}{MNRAS \textbf{524},  3692 (2023)} [\href{https://arxiv.org/abs/2307.02024}{arXiv:2307.02024}].
	
	\bibitem[Corcho et al.(2023b)]{corcho2023ageingb}	
	P. Corcho-Caballero, Y. Ascasibar, S. Sanchez, Á. López-Sánchez, \emph{Ageing and quenching through the ageing diagram: predictions from simulations and observational constraints}, \href{ https://doi.org/10.1093/mnras/stad147}{MNRAS \textbf{520},  193 (2023)} [\href{https://arxiv.org/abs/2208.14084}{arXiv:2208.14084}].
	
	\bibitem{fitts2017fire}	
	A. Fitts, M. Boylan-Kolchin, O. Elbert, J. Bullock, et al., \emph{Fire in the field: simulating the threshold of galaxy formation}, \href{ https://doi.org/10.1093/mnras/stx1757}{MNRAS \textbf{471},  3547 (2017)}.
	
	\bibitem{gensior2020heart}	
	J. Gensior, J. Kruijssen, B. Keller, \emph{Heart of darkness: the influence of galactic dynamics on quenching star formation in galaxy spheroids}, \href{https://doi.org/10.1093/mnras/staa1184}{MNRAS \textbf{495},  199 (2020)}.
	
	\bibitem{cortese2021dawes}	
	L .Cortese, .B Catinella, R. Smith, \emph{The Dawes Review 9: The role of cold gas stripping on the star formation quenching of satellite galaxies}, \href{https://doi.org/10.1017/pasa.2021.18}{PASA \textbf{38},  e035 (2021)}.
	
	\bibitem{brown2017cold}	
	T. Brown, B. Catinella, L. Cortese, C. Lagos, et al.,  \emph{Cold gas stripping in satellite galaxies: from pairs to clusters}, \href{ https://doi.org/10.1093/mnras/stw2991}{MNRAS \textbf{466},  1275 (2017)}.
	
	\bibitem{thorp2022almaquest}	
	M. Thorp, S. Ellison, H. Pan, L. Lin, et al., \emph{The ALMaQUEST Survey X: what powers merger induced star formation?}, \href{https://doi.org/10.1093/mnras/stac2288}{ApJ \textbf{516},  1462 (2022)}.
	
	\bibitem{akins2022quenching}	
	H. Akins, D. Narayanan, K. Whitaker, R. Davé, et al., \emph{Quenching and the UVJ Diagram in the SIMBA Cosmological Simulation}, \href{https://doi.org/10.3847/1538-4357/ac5d3a}{ApJ \textbf{929},  94 (2022)}.
	
	\bibitem{tacchella2022fast}	
	S. Tacchella, C. Conroy, S. Faber, B. Johnson, et al., \emph{Fast, slow, early, late: quenching massive galaxies at z$\sim$ 0.8}, \href{https://doi.org/10.3847/1538-4357/ac449b}{ApJ \textbf{926},  134 (2022)}.
	
	\bibitem{suess2022studying}	
	K. Suess, M. Kriek, R. Bezanson, J. Greene, et al., \emph{Studying Quenching in Intermediate-z Galaxies—Gas, Momentum, and Evolution}, \href{https://doi.org/10.3847/1538-4357/ac404a}{ApJ \textbf{926},  86 (2022)}.
	
	\bibitem[Belfiorel et al.(2018)]{belfiore2018sdss}
	F. Belfiore, R. Maiolino, K. Bundy, K. Masters, et al., \emph{SDSS IV MaNGA – sSFR profiles and the slow quenching of discs in green valley galaxies}, \href{https://doi.org/10.1093/mnras/sty768}{MNRAS  \textbf{477}, 3014(2018)}.
	
	\bibitem[Abdurro'uf et al.(2022)]{abdurro2022seventeenth}
	N. Abdurro'uf, K. Accetta, C. Aerts, V. Silva, et al., \emph{The Seventeenth Data Release of the Sloan Digital Sky Surveys: Complete Release of MaNGA, MaStar, and APOGEE-2 Data}, \href{https://doi.org/10.3847/1538-4365/ac4414}{ApJ  \textbf{259}, 2 (2022)}.
	
	\bibitem[Erfanianfar et al.(2016)]{erfanianfar2016non}
	G. Erfanianfar, P. Popesso, A. Finoguenov, D. Wilman, et al., \emph{Non-linearity and environmental dependence of the star-forming galaxies main sequence
	}, \href{https://doi.org/10.1093/mnras/stv2485}{MNRAS \textbf{455}, 2839 (2016)}[\href{https://arxiv.org/abs/1511.01899}{arXiv:1511.01899}].	
	
	\bibitem[Lang et al.(2014)]{lang2014bulge}
	P. Lang, S. Wuyts, R. Somerville, N. Schreiber, 
	R. Genzel, et al., \emph{Bulge Growth and Quenching since $z= 2.5$ in CANDELS/3D-HST
	}, \href{https://doi.org/10.1088/0004-637X/788/1/11}{ApJ \textbf{788}, 788 (2014)}[\href{https://arxiv.org/abs/1402.0866}{arXiv:1402.0866}].	
	
	\bibitem[Bluck et al.(2020)]{bluck2020galactic}
	A. Bluck, L. Maiolino, S. Sánchez, F. Sebastian, L. Sara, et al., \emph{ Are galactic star formation and quenching governed by local, global, or environmental phenomena?}, \href{https://doi.org/10.1093/mnras/stz3264}{MNRAS \textbf{492}, 96 (2020)}.
	
	\bibitem[Leja et al.(2022)]{leja2022new}
	J. Leja, S. Joshua, Y. Speagle, T. Benjamin, D. Johnson, et al., \emph{ A New Census of the $0.2 < z < 3.0$ Universe. II. The Star-forming Sequence}, \href{https://doi.org/10.3847/1538-4357/ac887d}{ApJ  \textbf{936}, 165 (2022)}.
	
	\bibitem[Thorne et al.(2020)]{thorne2021deep}
	J. Thorne, A. Robotham, L. Davies, S. Bellstedt, et al., \emph{Deep Extragalactic VIsible Legacy Survey (DEVILS): SED fitting in the D10-COSMOS field and the evolution of the stellar mass function and SFR--M$\star$ relation}, \href{https://doi.org/10.1093/mnras/stab1294}{MNRAS \textbf{505}, 540 (2021)}.
	
	\bibitem[Leslie et al.(2020)]{leslie2020vla}
	S. Leslie, E. Schinnerer, D. Liu, B. Magnelli, et al., \emph{ The VLA-COSMOS 3 GHz Large Project: Evolution of Specific Star Formation Rates out to $z \sim 5$}, \href{https://doi.org/110.3847/1538-4357/aba044}{ApJ \textbf{899}, 58 (2020)}.
	
	\bibitem[Croton et al.(2006)]{croton2006many}
	D. Croton, V. Springel, S. White, G. Lucia, et al., \emph{The many lives of active galactic nuclei: cooling flows, black holes and the luminosities and colours of galaxies }, \href{https://doi.org/10.1111/j.1365-2966.2005.09675.x}{MNRAS \textbf{365}, 11 (2013)} [\href{https://arxiv.org/abs/astro-ph/0508046}{arXiv:astro-ph/0508046}].
	
	\bibitem{rosario2012mean}
	D. Rosario, P. Santini, D. Lutz, L. Shao, et al., \emph{The mean star formation rate of X-ray selected active galaxies and its evolution from $z \sim 2.5$: results from PEP-Herschel}, \href{	https://doi.org/10.1051/0004-6361/201219258}{A\&A \textbf{545}, A45 (2012)} [\href{https://arxiv.org/abs/1203.6069}{arXiv:1203.6069}].
	
	\bibitem[Oemle et al.(2017)]{oemler2017star}
	A. Oemler, L. Abramson, M. Gladders, A.Dressler, et al., \emph{The Star Formation Histories of Disk Galaxies: The Live, the Dead, and the Undead}, \href{https://doi.org/10.3847/1538-4357/aa789e}{ApJ  \textbf{844}, 45 (2017)}.
	
	\bibitem[Croom et al.(2012)]{croom2012sydney}
	S. Croom, J. Lawrence, J. Bland-Hawthorn, J. Bryant et al., \emph{ The Sydney-AAO multi-object integral field spectrograph}, \href{https://doi.org/10.1111/j.1365-2966.2011.20365.x}{MNRAS \textbf{421}, 872 (2012)}.
	
	\bibitem[Bryant et al.(2005)]{bryant2015sami}
	J. Bryant, M. Owers, A. Robotham, S. Croom, et al., \emph{The SAMI Galaxy Survey: instrument specification and target selection}, \href{https://doi.org/10.1093/mnras/stu2635}{MNRAS \textbf{447}, 2857 (2005)}.
	
	\bibitem[Tempel et al.(2017)]{tempel2017merging}
	E. Tempel, T. Tuvikene, \emph{Merging groups and clusters of galaxies from the SDSS data. The catalogue of groups and potentially merging systems}, \href{https://doi.org/10.1051/0004-6361/201730499}{A\&A \textbf{602}, A100 (2017)} [\href{https://arxiv.org/abs/1704.04477}{arXiv:1704.04477}]. 
	
	\bibitem[Ade et al.(2016)]{collaborartion2016planck}
	P. Ade, N. Aghanim,  \emph{Planck 2015 results. XXIII. The thermal Sunyaev-Zeldovich effect--cosmic infrared background correlation}, \href{https://doi.org/10.1051/0004-6361/201527418}{A\&A \textbf{594}, A23 (2016)} [\href{https://arxiv.org/abs/1509.06555}{arXiv:1509.06555}].
	
	\bibitem[Eisenstein et al.(2011)]{eisenstein2011sdss}
	D. Eisenstein, D. Weinberg, \emph{SDSS-III: Massive spectroscopic surveys of the distant universe, the Milky Way, and extra-solar planetary systems}, \href{https://doi.org/10.1088/0004-6256/142/3/72}{AJ \textbf{142}, 72 (2011)} [\href{https://arxiv.org/abs/1101.1529}{arXiv:1101.1529}].
	
	\bibitem[Alam et al.(2015)]{alam2015eleventh}
	S. Alam, F. Albareti, C. Prieto, \emph{The Eleventh and Twelfth Data Releases of the Sloan Digital Sky Survey: Final Data from SDSS-III}, \href{https://doi.org/10.1088/0067-0049/219/1/12}{ApJS \textbf{219}, 12 (2015)} [\href{https://arxiv.org/abs/1501.00963}{arXiv:1501.00963}].
	
	\bibitem[Schlege et al.(1998)]{schlegel1998maps}
	D. Schlegel, D. Finkbeiner, M. Davis, \emph{ Maps of dust infrared emission for use in estimation of reddening and cosmic microwave background radiation foregrounds}, \href{https://doi.org/10.1086/305772}{ApJ \textbf{500}, 525 (1998)}.
	
	\bibitem[Schlege et al.(1998)]{strauss2002spectroscopic}
	M.  Strauss, D. Weinberg, R. Lupton, V. Narayanan, et al., \emph{ Spectroscopic target selection in the Sloan Digital Sky Survey: the main galaxy sample}, \href{https://doi.org/10.1086/342343}{AJ \textbf{124}, 3 (2002)}.
	
	\bibitem[Blanton et al.(2007)]{blanton2007k}
	M. Blanton, S. Roweis, \emph{K-corrections and filter transformations in the ultraviolet, optical, and near-infrared}, \href{https://doi.org/10.1086/510127}{AJ \textbf{133}, 734 (2007)} [\href{https://arxiv.org/abs/astro-ph/0606170}{arXiv:astro-ph/0606170}].
	
	\bibitem[Blanton et al.(2003)]{blanton2003galaxy}
	M. Blanton, D. Hogg, N. Bahcall, J. Brinkmann, et al., \emph{The galaxy luminosity function and luminosity density at redshift $z= 0.1$}, \href{https://doi.org/ 10.1086/375776}{AJ \textbf{592}, 819 (2003)} [\href{https://arxiv.org/abs/astro-ph/0210215}{arXiv:astro-ph/0210215}].
	
	\bibitem[Tempel et al.(2012)]{tempel2012groups}
	E. Tempel, E. Tago, L. Liivamägi, \emph{Groups and clusters of galaxies in the SDSS DR8-Value-added catalogues}, \href{https://doi.org/10.1051/0004-6361/201118687}{A\&A \textbf{540}, A106 (2012)}   [\href{https://arxiv.org/abs/1112.4648}{arXiv:1112.4648}].
	
	\bibitem[Tempel et al.(2014)]{tempel2014flux}
	E. Tempel, A. Tamm, M. Gramann, \emph{Flux- and volume-limited groups/clusters for the SDSS galaxies: catalogues and mass estimation}, \href{https://doi.org/10.1051/0004-6361/201423585}{A\&A \textbf{566},  A1 (2014)}   [\href{https://arxiv.org/abs/1402.1350}{arXiv:1402.1350}].	
	
	\bibitem[Ball et al.(2006)]{ball2006bivariatec}
	N. Ball, J. Loveday, R. Brunner, I. Baldry, J. Brinkmann, et al., \emph{ Bivariate galaxy luminosity functions in the Sloan Digital Sky Survey}, \href{https:// https://doi.org/10.1111/j.1365-2966.2006.11082.x}{MNRAS \textbf{373}, 845 (2006)}.
	
	\bibitem{schechter1976analytic}
	P. Schechter, \emph{ An analytic expression for the luminosity function for galaxies},  \href{https://ui.adsabs.harvard.edu/link_gateway/1976ApJ...203..297S/doi:10.1086/154079} {ApJ \textbf{203}, 297 (1976)}.
	
	\bibitem{deng2012some}
	X. Deng, Y. Xin, P. Wu, P. Jiang, et al., \emph{Some Properties of Active Galactic Nuclei in the Volume-limited Main Galaxy Samples of SDSS DR8}, \href{https://doi.org/10.1088/0004-637X/754/2/82}{ApJ \textbf{754}, 82 (2012)}.
	
	\bibitem[Kennicutt(1998)]{kennicutt1998star}
	C. Kennicutt, \emph{Star Formation in Galaxies Along the Hubble Sequence}, \href{
		https://doi.org/10.48550/arXiv.astro-ph/9807187
	}{A\&A \textbf{39},  189 (1998)}   [\href{https://arxiv.org/abs/astro-ph/9807187}{arXiv:astro-ph/9807187}].
	
	\bibitem[Salim et al.(2015)]{salim2015mass}
	S. Salim, J. Lee, R. Davé,  \emph{On the Mass-Metallicity-Star Formation Rate Relation for Galaxies at $ z\sim 2$}, \href{https://doi.org/10.1088/0004-637X/808/1/25}{ApJ \textbf{808}, 14pp (2015)} [\href{https://arxiv.org/abs/1506.03080}{arXiv:1506.03080}].
	
	\bibitem[Sanchez et al.(2018)]{sanchez2018sdss}
	S. Sanchez, V. Avila-Reese, H. Hernandez-Toledo, E. Cortes-Suarez, et al., \emph{ SSDSS IV MaNGA - Properties of AGN host galaxies}, \href{https://doi.org/10.48550/arXiv.1709.05438}{RxMAA \textbf{54}, 1 (2018)}.
	
	\bibitem[Fernandes et al.(2011)]{cid2011comprehensive}
	R. Cid-Fernandes, G. Stasinska, A. Mateus, N. Vale-Asari, \emph{A comprehensive classification of galaxies in the Sloan Digital Sky Survey: how to tell true from fake AGN}, \href{https://doi.org/0.1111/j.1365-2966.2011.18244.x7}{MNRAS \textbf{413}, 1687 (2011)}.
	
	\bibitem[Kewley et al.(2006)]{kewley2006host}
	L. Kewley, B. Groves, G. Kauffmann, T. Heckman, et al., \emph{The host galaxies and classification of active galactic nuclei}, \href{https://doi.org/10.1111/j.1365-2966.2006.10859.x}{MNRAS \textbf{372}, 961 (2006)} [\href{https://arxiv.org/abs/astro-ph/0605681}{arXiv:astro-ph/0605681}].
	
	\bibitem[Kauffmann et al.(2003)]{kauffmann2003host}
	G. Kauffmann, T. Heckman, C. Tremonti, J. Brinchmann, et al., \emph{The host galaxies of active galactic nuclei}, \href{https://doi.org/10.1111/j.1365-2966.2003.07154.x}{MNRAS \textbf{346}, 1055 (2003)} [\href{https://arxiv.org/abs/astro-ph/0304239}{arXiv:astro-ph/0304239}].
	
	\bibitem[Kewley et al.(2001)]{kewley2001theoretical}
	L. Kewley, M. Dopita, R. Sutherland, C. Heisler, et al., \emph{Theoretical modeling of starburst galaxies}, \href{https://doi.org/10.1086/321545}{ApJ \textbf{556}, 121 (2001)} [\href{https://arxiv.org/abs/astro-ph/0106324}{arXiv:astro-ph/0106324}].
	
	\bibitem[Singh et al.(2013)]{singh2013nature}
	R. Singh, G. van de Ven, K. Jahnke, M. Lyubenova, et al., \emph{The nature of LINER galaxies}, \href{https://doi.org/10.1051/0004-6361/201322062}{A\&A \textbf{558}, A43 (2013)}.
	
	\bibitem[Sanchez (2020)]{sanchez2020spatially}
	S. Sánchez, \emph{Spatially Resolved Spectroscopic Properties of Low-Redshift Star-Forming Galaxies}, \href{https://doi.org/10.1146/annurev-astro-012120-013326}{ARA\&A \textbf{99}, 58 (2020)} [\href{https://arxiv.org/abs/1911.06925}{arXiv:1911.06925}].
	
	\bibitem[Sanchez et al.(2021)]{sanchez2021local}
	S. Sánchez, C. Walcher, C. Lopez-Cobá, J. Barrera-Ballesteros, \emph{From Global to Spatially Resolved in Low-Redshift Galaxies}, \href{https://10.22201/ia.01851101p.2021.57.01.01}{RxMAA \textbf{57}, 3 (2021)} [\href{https://arxiv.org/abs/2009.00424}{arXiv:2009.00424}].
	
	\bibitem[Peng et al.(2015)]{peng2015strangulation}
	Y. Peng, R. Maiolino, R. Cochrane, \emph{Strangulation as the primary mechanism for shutting down star formation in galaxies}, \href{https://doi.org/10.1038/nature14439}{Nature \textbf{521}, 192 (2015)} [\href{https://arxiv.org/abs/1505.03143}{arXiv:1505.03143}].
	
	\bibitem[Mao et al.(2022)]{mao2022revealing}
	Z. Mao, T. Kodama, J. Pérez-Martínez, T. Suzuki, et al., \emph{Revealing impacts of stellar mass and environment on galaxy quenching}, \href{https://doi.org/10.1051/0004-6361/202243733}{A\&A \textbf{666}, A141 (2022)} [\href{https://arxiv.org/abs/2208.00722}{arXiv:2208.00722}].
	
	\bibitem[Rodríguez-Montero et al.(2019)]{rodriguez2019mergers}
	F. Rodríguez-Montero, R. Davé, V. Wild, D. Anglés-Alcázar, et al., \emph{Mergers, starbursts, and quenching in the simba simulation}, \href{https://doi.org/10.1093/mnras/stz2580}{MNRAS \textbf{490}, 2139 (2019)} [\href{https://arxiv.org/abs/1907.12680}{arXiv:1907.12680}].
	
	\bibitem[Tan et al.(2022)]{tan2022resolved}
	V. Tan, A. Muzzin, Z. Marsan, V. Sok, et al., \emph{Resolved Stellar Mass Maps of Galaxies in the Hubble Frontier Fields: Evidence for Mass Dependency in Environmental Quenching}, \href{https://doi.org/10.3847/1538-4357/ac7051}{ApJ \textbf{933}, 30 (2022)} [\href{https://arxiv.org/abs/2205.07913}{arXiv:2205.0791}].
	
	\bibitem[Ge et al.(2020)]{ge2020conditions}
	X. Ge, F. Liu, Q. Gu, E. Contini, et al., \emph{Conditions for galaxy quenching at $0.5 < z < 2.5$ from CANDELS: compact cores and environment}, \href{https://doi.org/10.1088/1674-4527/20/8/116}{RAA \textbf{20}, 116 (2022)}.
	
	\bibitem[Reeves, et al.(2021)]{reeves2021gogreen}
	A. Reeves, M. Balogh, R. Van der Burg, A. Finoguenov, et al., \emph{The GOGREEN survey: dependence of galaxy properties on halo mass at $z > 1 $ and implications for environmental quenching}, \href{https://doi.org/10.1093/mnras/stab1955}{MNRAS \textbf{506}, 3364 (2021)} [\href{https://arxiv.org/abs/2107.03425}{arXiv:2107.03425}].
	
	\bibitem[Corcho et al.(2021)]{corcho2021galaxy}
	P. Corcho-Caballero, J. Casado, Y. Ascasibar, R. García-Benito, et al., \emph{Galaxy evolution on resolved scales: ageing and quenching in CALIFA}, \href{https://doi.org/10.1093/mnras/stab2503}{MNRAS \textbf{507}, 5477 (2021)} [\href{https://arxiv.org/abs/2107.13478}{arXiv:2107.13478}].
	
	\bibitem[Hasan et al.(2023)]{hasan2023evolving}
	F. Hasan, J. Burchett, A. Abeyta, D. Hellinger, et al., \emph{The Evolving Effect of Cosmic Web Environment on Galaxy Quenching}, \href{https://doi.org/10.1093/mnras/stab2503}{MNRAS \textbf{950}, 114 (2023)} [\href{https://arxiv.org/abs/2303.08088}{arXiv:2303.08088}].
	
	\bibitem[Kipper et al.(2021)]{kipper2021role}
	R. Kipper, A. Tamm, E. Tempel, R. Propris, et al., \emph{The role of stochastic and smooth processes in regulating galaxy quenching}, \href{https://doi.org/10.1051/0004-6361/202039648}{A\&A \textbf{647}, A32 (2021)} [\href{https://arxiv.org/abs/2101.08549}{arXiv:2101.08549}].
	
	\bibitem[Einasto et al.(2022)]{einasto2022death}
	M. Einasto, R. Kipper, P. Tenjes, J. Einasto, et al., \emph{Death at watersheds: Galaxy quenching in low-density environments}, \href{https://doi.org/10.1051/0004-6361/202244304}{A\&A \textbf{668}, A69 (2022)}. [\href{https://arxiv.org/abs/2210.10761}{arXiv:2210.10761}].
	
	\bibitem[McElwain et al.(2023)]{mcelwain2023james}
	M. McElwain, L. Feinberg, M. Perrin, M. Clampin, et al., \emph{The James Webb Space Telescope Mission: Optical Telescope Element Design, Development, and Performance}, \href{https://doi.org/10.1088/1538-3873/acada0}{PASP \textbf{135}, 058001 (2023)}. [\href{https://arxiv.org/abs/2301.01779}{arXiv:2301.01779}].
	
    \bibitem[Drory et al.(2015)]{drory2015manga}	
    N. Drory, N. MacDonald, MA. Bershady, K. Bundy, et al., \emph{The MaNGA integral field unit fiber feed system for the Sloan 2.5 m telescope}, \href{ https://doi.org/10.1088/0004-6256/149/2/77}{AJ \textbf{149},  77 (2015)} [\href{https://arxiv.org/abs/1412.1535}{arXiv:1412.1535}].	
	
\end{thebibliography}
\end{document}